\documentclass[letterpaper,twocolumn,10pt]{article}
\PassOptionsToPackage{table,dvipsnames}{xcolor}
\usepackage{usenix-2020-09}
\usepackage[normalem]{ulem}
\usepackage{graphicx}

\usepackage{tikz}
\usetikzlibrary{arrows}
\usepackage{enumerate}
\usepackage{enumitem}
\usepackage{comment}
\usepackage{pifont}

\usepackage{booktabs}
\usepackage{tabularx}
\usepackage{makecell}
\usepackage{amssymb}
\usepackage{caption}
\usepackage{subcaption}
\usepackage[normalem]{ulem}
\usepackage{siunitx}

\usepackage{amsmath}
\usepackage[ruled,vlined,linesnumbered]{algorithm2e}

\usepackage[many]{tcolorbox}    	
\usepackage{multicol}               

\usepackage{booktabs}
\usepackage{multirow}

\usepackage{listings}
\usepackage[scaled=0.85]{beramono}

\usepackage{todonotes}

\newcommand{\insight}[1]{\colorbox{gray!15}{\textbf{#1}}}
\newcommand{\opportunity}[1]{\colorbox{blue!5}{\textbf{#1}}}
\definecolor{codegreen}{rgb}{0,0.6,0}
\definecolor{codegray}{rgb}{0.5,0.5,0.5}
\definecolor{codepurple}{rgb}{0.58,0,0.82}
\definecolor{backcolour}{rgb}{0.95,0.95,0.92}
\definecolor{textblue}{rgb}{.2,.2,.7}
\definecolor{textred}{rgb}{0.54,0,0}
\definecolor{textgreen}{rgb}{0,0.43,0}
\definecolor{codered}{rgb}{201,72,12}
\definecolor{codeblue}{rgb}{0.0, 0.0, 1.0}
\definecolor{awesome}{rgb}{1.0, 0.13, 0.32}
\definecolor{main}{HTML}{283618}    
\definecolor{sub}{HTML}{FBF8CC}     

\definecolor{mainI}{HTML}{283618}    
\definecolor{subI}{HTML}{FFCFD2}     

\newtcolorbox{boxK}{
    sharpish corners, 
    boxrule = 0pt,
    toprule = 4.5pt, 
    enhanced,
    fuzzy shadow = {0pt}{-2pt}{-0.5pt}{0.5pt}{black!35} 
}

\newtcolorbox{boxG}{
    enhanced,
    boxrule = 0pt,
    colback = sub,
    borderline west = {1pt}{0pt}{main}, 
    borderline west = {0.75pt}{2pt}{main}, 
    borderline east = {1pt}{0pt}{main}, 
    borderline east = {0.75pt}{2pt}{main}
}

\newtcolorbox{boxI}{
    enhanced,
    boxrule = 0pt,
    colback = subI,
    borderline west = {1pt}{0pt}{mainI}, 
    borderline west = {0.75pt}{2pt}{mainI}, 
    borderline east = {1pt}{0pt}{mainI}, 
    borderline east = {0.75pt}{2pt}{mainI}
}

\newcommand*{\Mname}{\textsc{GenServe}}

\newcommand{\dcircle}[2][0.4]{%
  \tikz[baseline=(char.base)]{
    \node[shape=circle, draw=black, fill=black,
          minimum size=2*#1 cm, inner sep=0pt] (char)
          {\textcolor{white}{#2}};
  }%
}

\begin{document}

\date{}

\title{\Large \bf \Mname{}: Efficient Co-Serving of Heterogeneous Diffusion Model Workloads}

\author{
{\rm Fanjiang Ye}$^{1}$ \quad
{\rm Zhangke Li}$^{1}$ \quad
{\rm Xinrui Zhong}$^{1}$ \quad
{\rm Ethan Ma}$^{1}$ \quad
{\rm Russell Chen}$^{1}$ \\
{\rm Kaijian Wang}$^{1}$ \quad
{\rm Jingwei Zuo}$^{1}$ \quad
{\rm Desen Sun}$^{2}$ \quad
{\rm Ye Cao}$^{3}$ \quad
{\rm Triston Cao}$^{4}$ \\
{\rm Myungjin Lee}$^{5}$ \quad
{\rm Arvind Krishnamurthy}$^{6}$ \quad
{\rm Yuke Wang}$^{1}$ \\[6pt]
$^{1}$Rice University \quad
$^{2}$University of Waterloo \quad
$^{3}$Independent Researcher \\
$^{4}$NVIDIA \quad
$^{5}$Cisco Research \quad
$^{6}$University of Washington
}

\maketitle

\begin{abstract}
Diffusion models have emerged as the prevailing approach for text-to-image (T2I) and text-to-video (T2V) generation, yet production platforms must increasingly serve both modalities on shared GPU clusters while meeting stringent latency SLOs.
Co-serving such heterogeneous workloads is challenging: T2I and T2V requests exhibit vastly different compute demands, parallelism characteristics, and latency requirements, leading to significant SLO violations in existing serving systems.
We present \Mname{}, a co-serving system that leverages the inherent predictability of the diffusion process to optimize serving efficiency.
A central insight is that diffusion inference proceeds in discrete, predictable steps and is naturally preemptible at step boundaries, opening a new design space for heterogeneity-aware resource management.
\Mname{} introduces step-level resource adaptation through three coordinated mechanisms: intelligent video preemption, elastic sequence parallelism with dynamic batching, and an SLO-aware scheduler that jointly optimizes resource allocation across all concurrent requests.
Experimental results show that \Mname{} improves the SLO attainment rate by up to 44\% over the strongest baseline across diverse configurations.

\end{abstract}

\section{Introduction}

Recently, diffusion models have achieved remarkable breakthroughs and widespread popularity~\cite{ho2020denoising, rombach2022high, dhariwal2021diffusion}, demonstrating unparalleled performance in generative tasks.
From U-Net architectures~\cite{song2020denoising, rombach2022high, ho2020denoising} to Diffusion Transformers (DiT)~\cite{peebles2023scalable, esser2024scaling, bao2023all}, these models have established themselves as the dominant paradigm for both text-to-image (T2I)~\cite{ramesh2022hierarchical, saharia2022photorealistic, podell2023sdxl} and text-to-video (T2V)~\cite{singer2022make, blattmann2023stable} generation. As demand grows, production platforms need to serve both modalities under tight latency service-level objectives (SLOs) on heterogeneous diffusion workloads.
%
%
In practice, serving image and video generation jointly on shared clusters is increasingly common in production platforms ~\cite{yin2026vllm, qiu2025modserve, ma2025cornserve}.
Users frequently submit both T2I and T2V requests to the same service endpoint, and the workload mix fluctuates over time.
Co-serving both modalities on a unified cluster is attractive because DiT models are compact enough to fit on a single GPU without model parallelism~\cite{fang2024xdit, lu2026tetriserve}, and T2I and T2V can share the same DiT backbone, allowing a shared deployment to achieve higher resource utilization than dedicating separate device pools to each modality.

However, co-serving mixed image and video workloads on shared clusters remains highly challenging. The core difficulty is the mismatch between heterogeneous workload demands and the static resource management of existing systems. T2I and T2V requests differ by orders of magnitude in per-step runtime cost (up to 20$\times$ for 720p Image and Video) and impose vastly different latency targets, yet current systems treat them with a single, fixed serving configuration.
%
Beyond the workload asymmetry, this heterogeneity interacts poorly with conventional scheduling policies. Under FIFO, a small fraction of long-running video requests causes severe head-of-line (HOL) blocking, starving latency-sensitive image requests. Under shortest-job-first (SJF), image requests are always prioritized, causing video starvation and missed video deadlines. Neither policy can balance the competing latency targets of both modalities, leading to widespread SLO violations.
Moreover, existing diffusion serving systems~\cite{fang2024xdit, ahmad2025diffserve} are designed for homogeneous workloads, applying a single, statically configured strategy for batching~\cite{crankshaw2017clipper, yu2022orca, ahmad2025diffserve} and parallelism~\cite{li2024distrifusion, fang2024pipefusion} that is tuned for one task type. These assumptions break down in mixed image and video serving, where no static strategy fits all requests: the optimal configuration varies continuously with the workload mix, request resolutions, and device state. 
To the best of our knowledge, no existing system supports co-serving heterogeneous diffusion workloads with coordinated resource management across task types.


\begin{figure}[t]
  \begin{center}
\includegraphics[width=\linewidth]{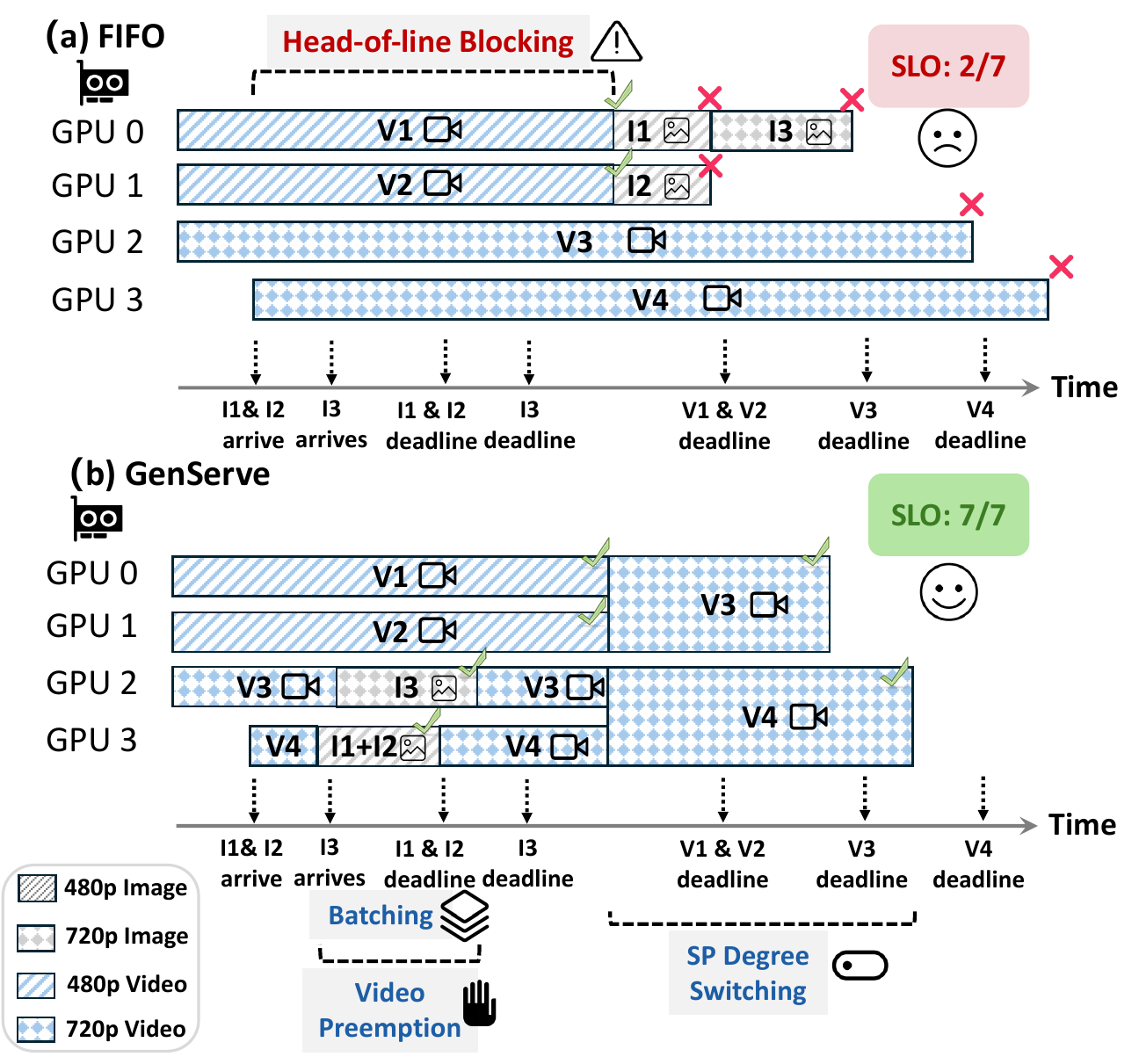}
  \end{center}
  \caption{Serving 4 videos (V1--V4) and 3 images (I1--I3) on 4 GPUs. (a)~FIFO: videos occupy all GPUs until completion; images must wait, causing HOL blocking and 5 deadline misses (SLO 2/7). (b)~\Mname{}: when I1, I2 arrive, the scheduler preempts V3 and V4 at step boundaries, batches I1+I2 on GPU~3, and serves I3 on GPU~2. After images complete, V3 and V4 resume with SP degree switching (each scaling to 2 GPUs) to recover lost slack. All 7 requests meet their deadlines (SLO 7/7).}
  \label{fig: sec_1_overall_process}
\end{figure}

To overcome these issues, we identify several key opportunities. 
\underline{First}, unlike LLM inference whose runtime depends on unpredictable output length~\cite{kwon2023efficient, yu2022orca}, diffusion models exhibit highly predictable execution: they perform a fixed number of denoising steps whose per-step cost is determined solely by input configuration (resolution and frame count). This predictability enables the system to accurately estimate the remaining cost of every in-flight request, making scheduling decisions controllable and resource-aware at step-level granularity.
%
\underline{Second}, current systems~\cite{zheng2024open, zheng2025open,fang2024xdit} execute the entire diffusion pipeline as an atomic unit, offering no mechanism to interrupt a long-running request once it begins. Yet diffusion models are naturally preemptible at step boundaries: a request can be paused after any denoising step and resumed later without losing progress or quality. The intermediate state is a compact latent tensor~\cite{rombach2022high, ye2025supergen} (typically up to tens of MBs, Table~\ref{tab:paused_memory}) that can be retained in GPU memory, making pause and resume a millisecond-level operation with negligible switching overhead (as measured in \S\ref{sec:eval}). Combined with predictable per-step costs, this enables fine-grained, deadline-aware preemption at near-zero cost.
\underline{Third}, the distinct computational profiles of T2I and T2V provide opportunities for workload-aware resource reallocation: dynamic batching can improve T2I throughput on underutilized GPUs, while elastic Sequence Parallelism (SP)~\cite{jacobs2023deepspeed, liu2023ring, fang2024xdit}, which partitions the input token sequence across GPUs to reduce per-step latency, can be dynamically adjusted to balance latency reduction against communication overhead for T2V tasks of varying resolutions. Rather than fighting heterogeneity with a one-size-fits-all policy, the system can exploit it by adapting resource allocation per request type at step-level runtime.

Based on the above insights, we propose \Mname{}, an efficient diffusion serving system for mixed image and video workloads that exploits the predictability of diffusion execution to enable heterogeneity-aware and step-level resource adaptation. The core idea of \Mname{} is to transform a reactive serving problem into a proactive optimization: because per-step costs are stable and known, the system can predict the impact of every scheduling action, including batching, parallelism, reconfiguration and preemption before committing to it. At each scheduling round, the system solves a lightweight allocation that jointly considers all requests and maximizes global SLO attainment.


\begin{figure}[t]
\small
\begin{lstlisting}[caption={Example usage of {\Mname{}}.}, label={lst:genserve-api}]
import GenServe

# Configure devices and models
server = GenServe.Server(
    GPUs="0, 1, 2, 3, ..., 8",
    image_model="stabilityai/stable-diffusion-3.5",
    video_model="Wan-AI/Wan2.2-T2V-5B"
)
# Set per-modality SLO targets (in seconds)
server.set_slo(image_slo=3.0, video_slo=60.0)
# Load offline latency profiles for the scheduler
server.load_profiler(profile_dir="profiles/")
# Enable serving optimizations
server.enable(
    preemption=True,          # Preemption
    elastic_sp=[1, 2, 4, 8],  # Allowed SP degrees
    dp_solver=True,           # SLO-aware DP scheduler
    batching=True,            # Dynamic image batching
)
# Load mixed request trace and launch serving
server.load_requests("traces/workload.json")
results = server.serve()
\end{lstlisting}
\vspace{-10pt}
\end{figure}


Specifically, \Mname{} integrates a set of complementary mechanisms that jointly address the challenges of heterogeneous diffusion serving (Figure~\ref{fig: sec_1_overall_process}). 
\underline{First}, we design an intelligent video preemption mechanism that leverages per-step predictability to determine when and which running video requests can be paused at step boundaries to release GPUs for urgent image requests, while ensuring that each preempted video's own deadline remains feasible upon resumption. 
\underline{Second}, we introduce elastic resource allocation that dynamically adjusts sequence parallelism degrees for video tasks and batching strategies for image tasks based on runtime workload characteristics, exploiting the distinct computational profiles of each modality rather than applying a uniform policy. 
\underline{Third}, we develop a stepwise SLO-aware scheduler that formulates joint resource allocation as a knapsack-style dynamic programming problem, selecting the combination of batching budgets, parallelism configurations, and preemption candidates that maximizes deadline satisfaction across all concurrent requests. 
To make these capabilities accessible, we expose a simple usage example (Listing~\ref{lst:genserve-api}) that allows developers to perform serving and easily extend other diffusion models.

To sum up, we make the following contributions:

\begin{itemize}
    \item We propose the intelligent video preemption (\S\ref{subsec:preemption}) and elastic resource allocation (\S\ref{subsec:Elastic-Resource-Reallocation}) that exploit the predictability and step-level preemptibility of diffusion models to dynamically adjust parallelism degrees and batching strategies at runtime.
    \item We design a stepwise SLO-aware scheduler (\S\ref{sec:slo-aware-scheduler}) that formulates joint resource allocation as a lightweight DP-based optimization to maximize deadline satisfaction across all concurrent requests.
    \item Extensive evaluation demonstrates that \Mname{} achieves up to 44\% improvement in SLO attainment over state-of-the-art baselines (\S\ref{sec:eval}).
\end{itemize}

\section{Background}
This section provides background on diffusion model inference and summarizes prior serving systems. Section~\ref{sec2: diffusion-models} introduces the diffusion generation pipeline and explains why mixed T2I/T2V serving on shared GPUs is challenging. Section~\ref{sec2: heterogeneous-serving} introduces existing diffusion serving optimizations and highlights the limitations on mixed workloads.

\subsection{Diffusion Models for Content Generation}
\label{sec2: diffusion-models}
Latent Diffusion Models (LDMs) are the dominant paradigm for both text-to-image (T2I) and text-to-video (T2V) generation. They operate in a compressed latent space where an encoder $\mathcal{E}$ maps input $x$ to $z{=}\mathcal{E}(x)$, and generation proceeds via forward (noising) and reverse (denoising) phases. The forward phase adds Gaussian noise over $T$ timesteps with schedule $\{\beta_t\}$:
\begin{equation} \small
\label{equation: forward diffusion} 
q(\mathbf{z}_t \mid \mathbf{z}_{t-1}) = \mathcal{N} \left( \mathbf{z}_t; \sqrt{1 - \beta_t} \, \mathbf{z}_{t-1}, \beta_t \mathbf{I} \right).
\end{equation}
The reverse phase iteratively recovers $z_0$ from $z_T$ via a learned network:
\begin{equation} \small
\label{equation: reverse process of diffusion}
    p_\theta(\mathbf{z}_{t-1} \mid \mathbf{z}_t) = \mathcal{N} \left( \mathbf{z}_{t-1}; \mu_\theta(\mathbf{z}_t, t), \Sigma_\theta(\mathbf{z}_t, t) \right).
\end{equation}

\begin{figure}[t]
  \begin{center}
\includegraphics[width=\linewidth]{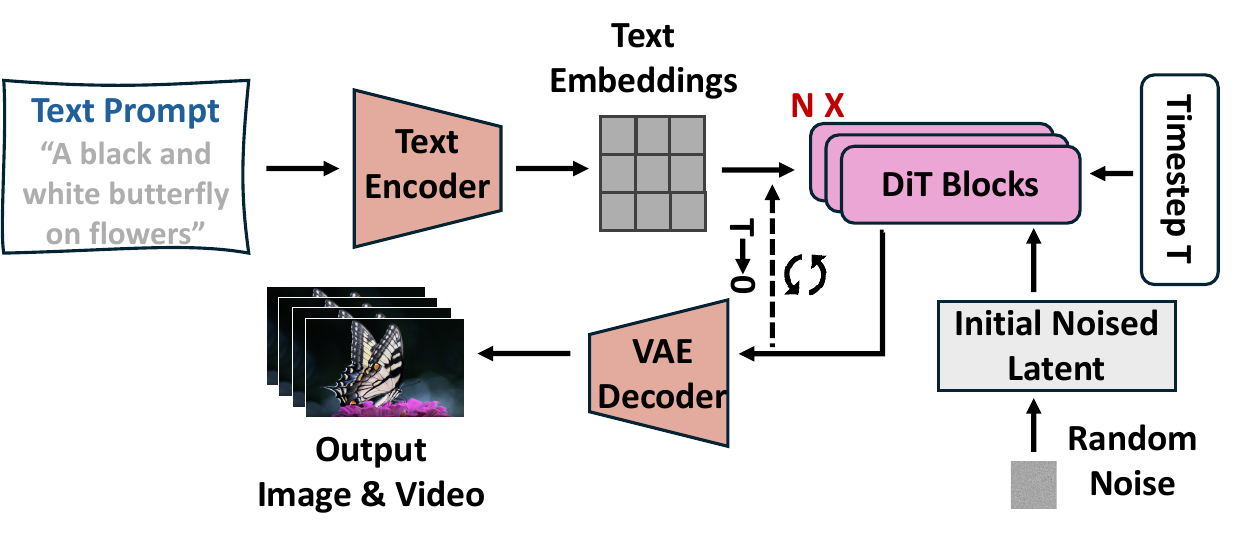}
  \end{center}
  \caption{Overview of DiT inference process.}
  \label{fig: sec_2_Diffusion-Process}
\end{figure}

As shown in Figure~\ref{fig: sec_2_Diffusion-Process}, modern Diffusion Transformer (DiT)~\cite{hong2022cogvideo, wan2025wan, kong2024hunyuanvideo} pipelines share three components across T2I and T2V: a VAE for latent encoding/decoding, a text encoder for prompt embeddings, and DiT blocks that perform $N$ denoising steps with cross-attention for text alignment. The critical difference lies in the latent structure: T2I operates on a 2D spatial grid, while T2V adds a temporal dimension proportional to the frame count, substantially increasing per-step token count and running latency. Combined with more denoising steps, a single video request can occupy multiple GPUs for tens of minutes, orders of magnitude longer than an image.

This asymmetry creates a core tension for mixed-workload serving. Unlike LLMs, whose KV caches can consume the majority of GPU memory, DiT models have a relatively compact memory footprint, e.g., a single GPU can host both a T2I and a T2V model weights simultaneously, with a peak memory around 50 Gigabytes.
This compact footprint makes co-serving both modalities on shared GPUs feasible, while naively isolating the two workloads into dedicated GPU pools would leave substantial memory and compute capacity underutilized.
However, co-serving without careful coordination leads to poor user experience: long-running video tasks monopolize GPU compute, starving latency-sensitive image requests, while unmanaged contention causes deadline violations on both sides.

\subsection{Diffusion Model Serving}

\label{sec2: heterogeneous-serving}
Recent work has advanced diffusion serving along three axes. On the \textit{parallelism} front, prior systems have combined sequence parallelism with patch-level pipelining~\cite{fang2024xdit}, disaggregated pipeline stages with per-request SP degree selection~\cite{huang2025ddit}, and independently scaled each stage to reduce inter-stage imbalance~\cite{xia2025tridentserve}. On the \textit{scheduling} front, recent work has explored step-level dynamic SP with round-based deadline-aware scheduling~\cite{lu2026tetriserve} and prompt routing through model cascades under dynamic resource allocation~\cite{ahmad2025diffserve}. Complementarily, caching approaches~\cite{agarwal2024approximate, xia2026modm, sun2026mixfusion} reduce per-request cost by reusing intermediate latents or final outputs for similar prompts, though these are inherently lossy, affecting the output quality.

Despite these advances, all existing systems target a single task type and assume homogeneous workloads. The previous \textit{parallelism} strategies are tuned for one workload profile, without accounting for the fundamentally different scaling behaviors of T2I and T2V, and the previous \textit{scheduling} algorithms optimize for a single latency objective rather than balancing competing SLO targets. To the best of our knowledge, no existing system supports co-serving heterogeneous diffusion workloads, which demands coordinated resource management across task types with diverse computational profiles and latency objectives.
\section{Motivation}
\label{sec: motivation}



\begin{table}[t]
    \footnotesize
    \centering
    \caption{Per-step DiT runtime stability across batch sizes and SP degrees. Std and CV (Coefficient of Variation) are measured over repeated 50-step runs.}
    \begin{tabular}{cc cc c cc cc}
    \toprule
    \multicolumn{4}{c}{Image (SD 3.5)} & & \multicolumn{4}{c}{Video (Wan2.2-5B)} \\
    \cmidrule{1-4} \cmidrule{6-9}
    BS & SP & Std (ms) & CV (\%) & & BS & SP & Std (ms) & CV (\%) \\
    \midrule
    1 & 1 & 0.03 & 0.04 & & 1 & 1 & 0.12 & 0.02 \\
    2 & 1 & 0.04 & 0.03 & & 1 & 2 & 0.10 & 0.01 \\
    4 & 1 & 0.05 & 0.02 & & 1 & 4 & 0.16 & 0.02 \\
    8 & 1 & 0.07 & 0.01 & & 1 & 8  & 0.14 & 0.01 \\
    \bottomrule
    \end{tabular}
    \label{tab:per_step_stability}
\end{table}

\noindent \insight{Insight 1}: Diffusion models have predictable and stable per-step runtime, yet heterogeneous workloads naturally cause HOL blocking, making step-level preemption both feasible and necessary.

\begin{table}[t]
    \footnotesize
    \centering
    \caption{Stage-level runtime breakdown of T2V generation (Wan2.2-5B, 81 frames) on a single GPU. The DiT denoising stage dominates across all resolutions (92--95\% of total time).}
    \resizebox{\linewidth}{!}{
    \begin{tabular}{lcccc}
    \toprule
    Resolution & Text Enc. (s) & DiT (s) & VAE Dec. (s) & DiT Ratio \\
    \midrule
    256p & 0.03  & 4.41   & 0.34  & 92.2\% \\
    480p & 0.03  & 16.03  & 1.01  & 93.9\% \\
    720p & 0.03  & 50.00  & 2.47  & 95.2\% \\
    \bottomrule
    \end{tabular}}
    \label{tab:stage_breakdown}
\end{table}
\begin{figure}[t]
  \begin{center}
\includegraphics[width=\linewidth]{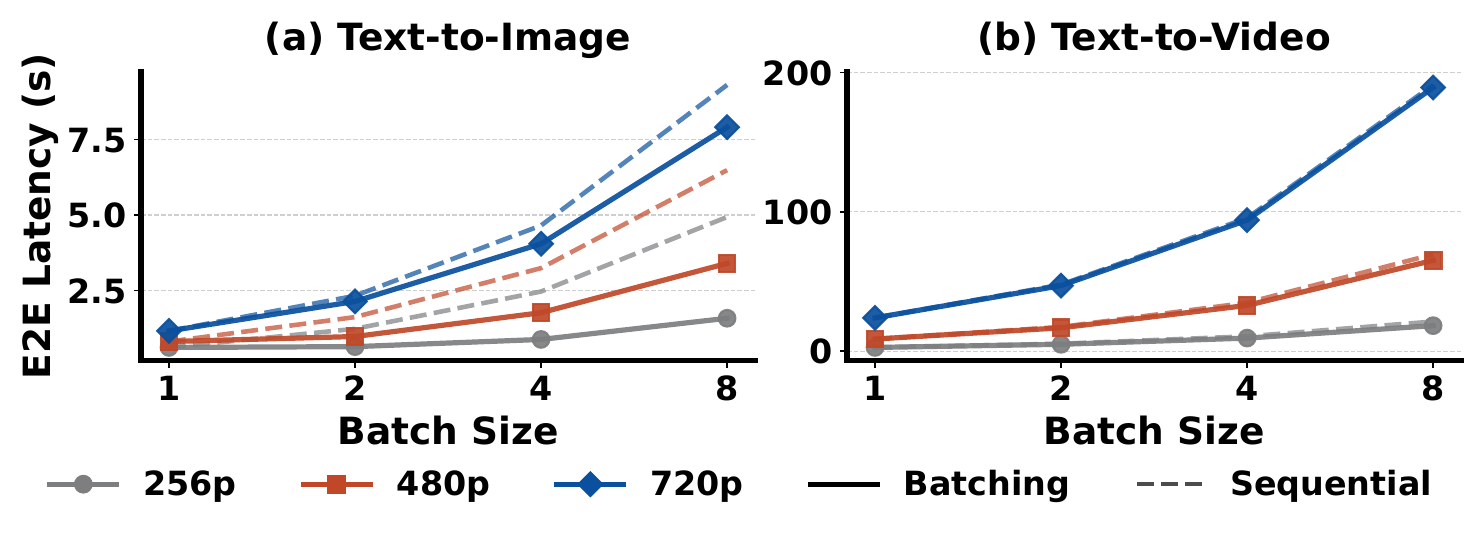}
  \end{center}
    \caption{End-to-end latency of T2I and T2V workloads across batch sizes and resolutions. In T2I, batching yields noticeable savings over theoretical sequential execution at low resolutions, while T2V exhibits limited room for batching-based latency reduction even for low-resolution requests.}
  \label{fig: sec_3_batchsize-effect-latency-throughput}
\end{figure}

Unlike LLM inference, whose runtime depends on the unpredictable length of generated output, diffusion models perform a fixed number of denoising steps whose per-step cost is determined by input configuration. 
As shown in Table~\ref{tab:per_step_stability}, the per-step runtime remains highly stable across batch sizes and SP degrees (CV $<$ 0.05\%). 
Meanwhile, as shown in Table~\ref{tab:stage_breakdown}, the DiT denoising stage occupies 92--95\% of the total generation time, and the process is inherently iterative, executing tens of independent steps in sequence. This predictability and iterative structure make it possible to accurately estimate the remaining cost of any request and to pause execution at step boundaries during the DiT stage without losing progress. 

\begin{figure}[t]
  \begin{center}
    \includegraphics[width=\linewidth]{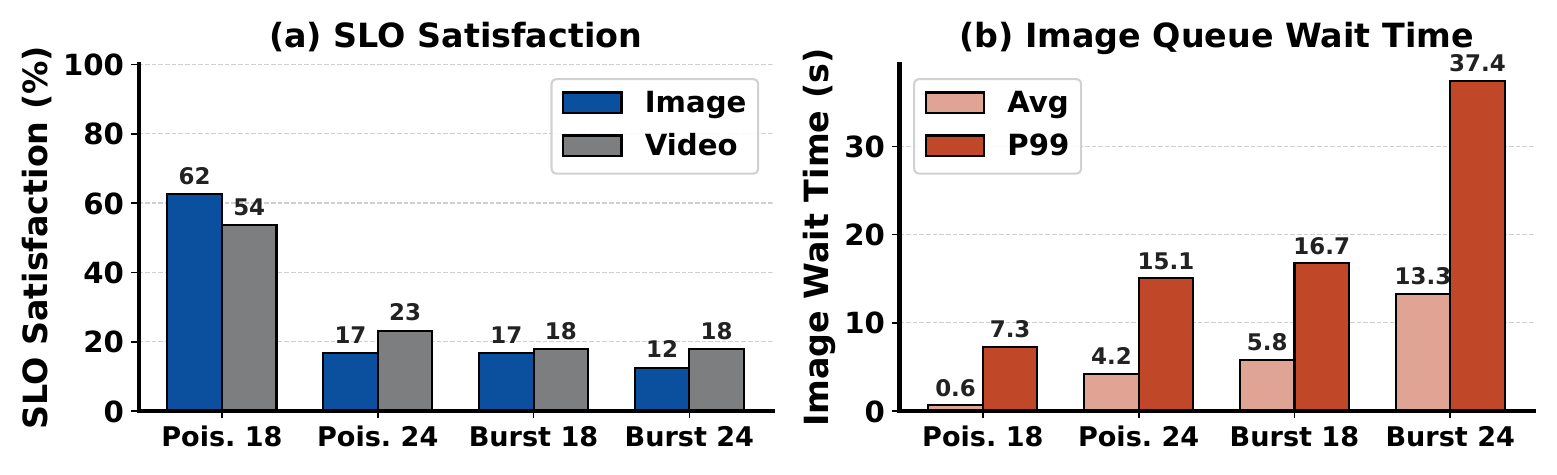}
  \end{center}
  \caption{Head-of-line blocking under FCFS scheduling with workloads (70\% video, 81 frames). Here Pois. means the Poisson arrival pattern. (a)~SLO satisfaction drops sharply under bursty video arrivals, image SLO falls from 62\% to 12\% as videos monopolize all GPUs. (b)~Image P99 queue wait time increases by $5\times$ from 7.3\,s to 37.4\,s, confirming that FCFS cannot protect image requests, causing HOL blocking.}
  \label{fig:sec_3_HOL-blocking}
\end{figure}

Furthermore, the compact memory footprint of DiT models (e.g., Wan2.2-T2V is only 5B) allows both sets of weights to reside in GPU memory simultaneously without swapping in and out GPUs, making time-based GPU sharing a natural deployment choice that fully utilizes device memory capacity and avoids costly weight migration. 
Under this co-location model, T2I and T2V requests share GPU compute time, and the order-of-magnitude runtime gap between them (Figure~\ref{fig: sec_3_batchsize-effect-latency-throughput}) makes the scheduling policy critical.
Conventional scheduling policies like First-in-First-out (FIFO) cause severe head-of-line blocking: as shown in Figure~\ref{fig:sec_3_HOL-blocking}, under bursty video arrivals, long-running video requests monopolize GPUs, causing image SLO satisfaction to drop from 62\% to 12\%.
Shortest-Job-First (SJF) always prioritizes image requests due to their much shorter execution time, starving video requests and causing widespread video SLO violations (as measured in \S\ref{sec:eval}).
None of these policies can balance the competing latency targets of both modalities.

\noindent {\opportunity{Opportunity 1}}: We can leverage the predictable, iterative nature of diffusion models to \textbf{preempt long-running video requests} at denoising step boundaries, releasing GPU compute for urgent image requests without wasting completed work or swapping model weights.


\noindent \insight{Insight 2}: T2I and T2V workloads exhibit fundamentally different compute profiles, requiring specific batching and parallelism strategies rather than a uniform configuration.

\begin{table}[t]
    \footnotesize
    \centering
    \caption{Per-step arithmetic intensity of DiT for T2I (SD3.5) and T2V (Wan2.2-5B, 81 frames) in BF16.}
    \resizebox{\linewidth}{!}{
    \begin{tabular}{llrrr}
    \toprule
    Model & Res. & Seq. Len. & FLOPs/step (T) & AI (FLOPs/B) \\
    \midrule
    \multirow{3}{*}{SD3.5 (T2I)}
     & 256p  & 256   & 0.36  & 243    \\
     & 480p  & 900   & 1.34  & 764    \\
     & 720p  & 2,304 & 3.91  & 1,646 \\
    \midrule
    \multirow{3}{*}{Wan2.2 (T2V)}
     & 256p  & 1,344   & 10.81  & 1,197   \\
     & 480p  & 4,725  & 43.90 & 3,437 \\
     & 720p  & 12,096 & 145.26 & 6,941 \\
    \bottomrule
    \end{tabular}}
    \label{tab:arithmetic_intensity}
\end{table}

%
For batching, we first estimate the per-step arithmetic intensity (AI) of DiT in Table~\ref{tab:arithmetic_intensity}, showing that T2V tends to be compute-intensive at all resolutions, meaning the DiT stage already saturates GPU compute even at batch size 1. Increasing the batch size, therefore, provides negligible throughput gain for T2V while worsening SLO attainment. In contrast, T2I at lower resolutions falls below the ridge point, leaving GPU utilization headroom that batching can exploit to improve runtime latency, as confirmed in Figure~\ref{fig: sec_3_batchsize-effect-latency-throughput}.

\begin{figure}[t]
  \begin{center}
\includegraphics[width=\linewidth]{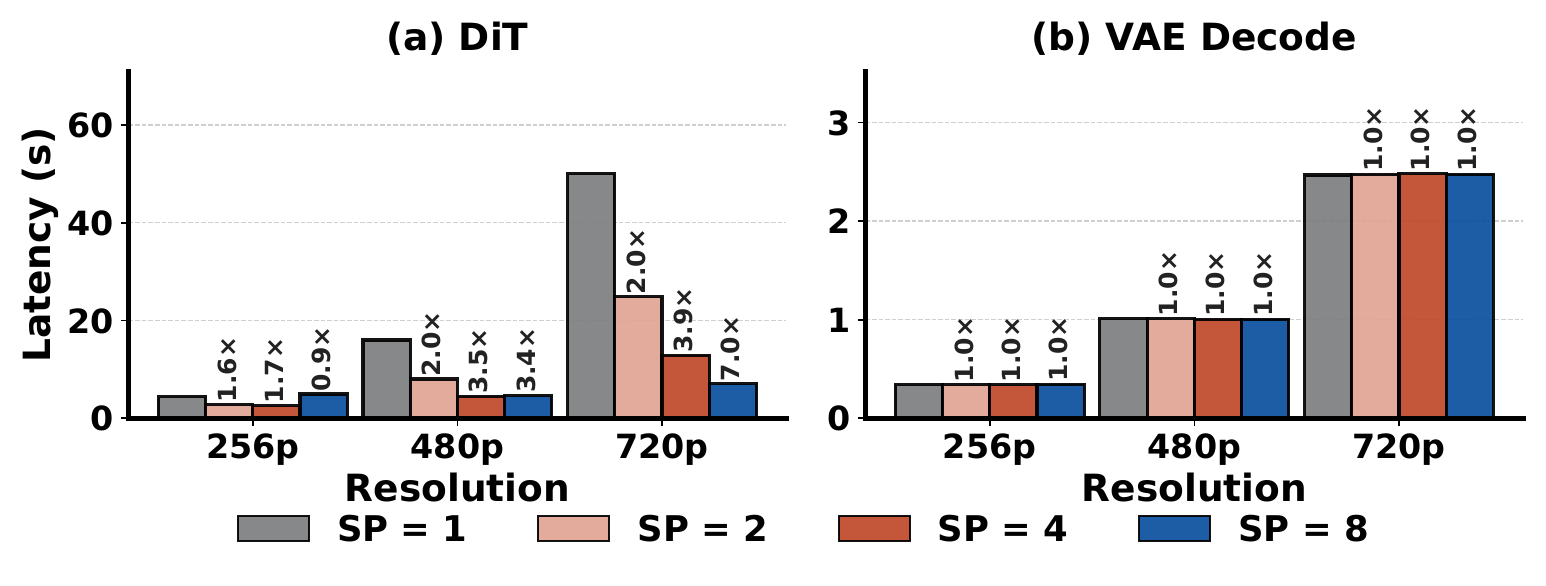}
  \end{center}
  \caption{The runtime of different stages in T2V across different resolutions and Sequence Parallelism (SP) degrees. DiT and VAE Decode latency of T2V across resolutions and SP degrees. DiT benefits from higher SP at high resolutions (up to 7.0$\times$ at 720p/81f) but shows diminishing returns at low resolutions. VAE Decode latency is unaffected by SP degree.}
  \label{fig: sec_3_video_stage_breakdown}
\end{figure}

\begin{figure}[t]
  \begin{center}
\includegraphics[width=\linewidth]{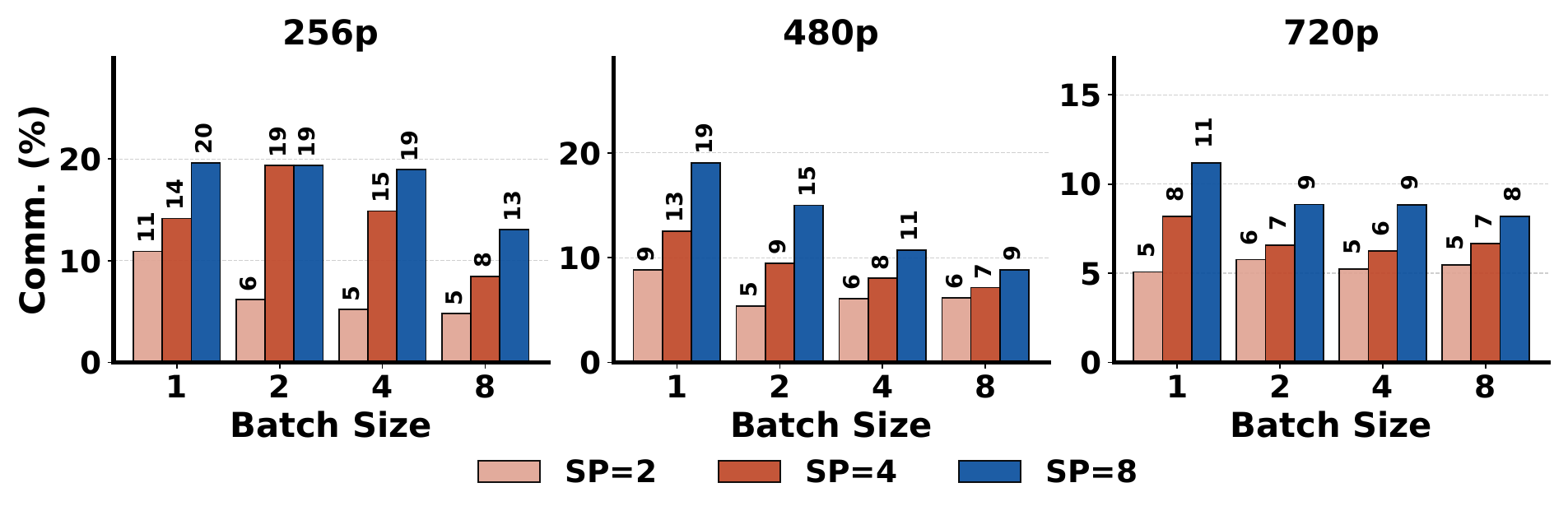}
  \end{center}
  \caption{Communication overhead (as \% of per-step time) for T2V DiT across resolutions, batch sizes, and SP degrees. Higher SP degrees incur more communication; at low resolutions (256p) the ratio reaches 20\%, explaining the poor SP scalability in Figure~\ref{fig: sec_3_video_stage_breakdown}. Larger batch sizes and higher resolutions reduce the ratio by increasing per-GPU computation.}
  \label{fig: sec_3_SP-degree-comm-comp-breakdown}
\end{figure}

%
For Sequence Parallelism, its effectiveness depends on whether the per-GPU computation is large enough to amortize the communication overhead. Figure~\ref{fig: sec_3_video_stage_breakdown} and Figure~\ref{fig: sec_3_SP-degree-comm-comp-breakdown} confirm this across both stages. \underline{For DiT}, high-resolution video generation provides sufficient per-GPU workload, so increasing SP yields near-linear latency reduction initially. However, as SP degree grows further, the communication fraction rises rapidly and per-GPU workload shrinks below efficient kernel granularity, causing the speedup to saturate. At low resolutions, this saturation occurs much earlier, making SP beyond a small degree ineffective. \underline{For VAE}, the computation is inherently too small to benefit from SP regardless of resolution: even under parallel execution~\cite{von-platen-etal-2022-diffusers}, GPUs operate over the same tensor, leading to low utilization and negligible speedup.

\noindent {\opportunity{Opportunity 2}}: We can \textbf{dynamically adapt resource allocation} per task type at runtime: applying aggressive batching to T2I for higher utilization, tuning SP degrees for T2V based on resolution, and keeping VAE on a single GPU.


\noindent \insight{Insight 3}: Static deployment and offline solvers cannot efficiently serve mixed diffusion workloads whose composition fluctuates at runtime, necessitating online scheduling that jointly considers all in-flight requests.

Existing systems~\cite{fang2024xdit} fix the SP degree for each request at admission time and never adjust it during execution. While this simplifies deployment, it cannot react to the GPU availability changes caused by request arrivals, completions, and preemptions in a mixed-workload setting. An alternative is to formulate optimal resource allocation as a Mixed-Integer Linear Program (MILP)~\cite{jain2001algorithms}, but MILP solvers are prohibitively slow for online use: even moderate problem sizes require seconds to minutes, far exceeding the millisecond-scale scheduling budget at step boundaries.

Beyond solver latency, the fundamental challenge is that the workload mix changes continuously during real serving: the ratio of image to video requests, their resolutions, and their remaining deadlines. A static partitioning of GPUs between T2I and T2V cannot track these dynamics, as over-provisioning one modality inevitably starves the other (as discussed in Figure~\ref{fig:e2e_E6_dedicated}). Meanwhile, preemption and adaptive resource allocation are individually beneficial, but applying them without joint coordination leads to suboptimal decisions: preemption releases GPUs that are allocated inefficiently, while per-request parallelism tuning without global visibility may improve one request at the expense of overall SLO attainment.

\noindent {\opportunity{Opportunity 3}}: We can combine per-step predictability with global request visibility to \textbf{formulate a lightweight online SLO-aware scheduler} that jointly decides preemption, batching, and parallelism at each scheduling round, maximizing SLO attainment across all concurrent requests under changing workload conditions.

\section{\Mname{}}
\label{sec:design}
\subsection{Overview}
\label{sec:overview}

Motivated by the observations in \S\ref{sec: motivation}, we present \Mname{}, an efficient serving system for mixed diffusion workloads that continuously reallocates GPU capacity across heterogeneous jobs while maintaining deadline visibility.
\Mname{} addresses the challenges through three mechanisms:
\emph{intelligent video preemption} (\S\ref{subsec:preemption}) that pauses running videos to release resources for urgent images without violating video deadlines;
\emph{elastic resource reallocation} (\S\ref{subsec:Elastic-Resource-Reallocation}) that adapts sequence parallelism degrees and batching images at runtime to increase the serving efficiency;
and a \emph{stepwise SLO-aware scheduler} (\S\ref{sec:slo-aware-scheduler}) that optimizes resource allocation across hybrid requests.


\begin{figure}[t]
  \centering
  \includegraphics[width=\linewidth]{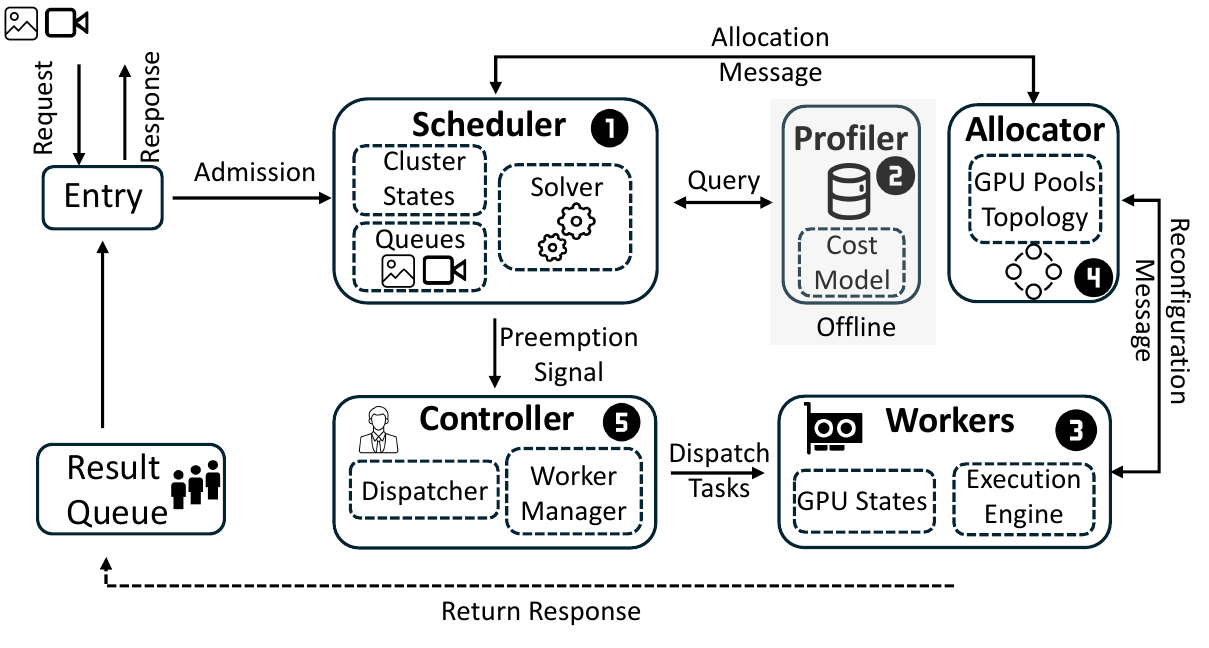}
  \caption{System overview of \Mname{}. 
  Incoming requests are admitted by the Scheduler, which maintains cluster states and queries the Profiler for offline latency estimates to invoke the Solver for joint optimization. The resulting decisions are passed to the Allocator and the Controller, which dispatches tasks and preemption signals to Workers. Workers execute inference on assigned GPUs and report step-level progress back through the Result Queue.}
  \label{fig:system-overview}
\end{figure}


Figure~\ref{fig:system-overview} illustrates the architecture of \Mname{}, consisting of five components that form a closed serving loop.
When a request arrives, the \underline{Scheduler}~(\dcircle[0.18]{1}) records its task type, resolution, and deadline, and maintains per-GPU state by processing worker events.
For images under low contention, the Scheduler consults the \underline{Profiler}~(\dcircle[0.18]{2}), which stores offline latency estimates across resolutions, batch sizes, and parallelism degrees, to coalesce same-resolution requests into a deadline-feasible batch and dispatches it to a \underline{Worker}~(\dcircle[0.18]{3}).
For videos, the Scheduler enqueues the request and triggers a scheduling round: the solver constructs a global snapshot, queries the Profiler for cost estimates, and solves a joint optimization over all in-flight videos and queued images.
If preemption or SP reconfiguration is needed, the \underline{Allocator}~(\dcircle[0.18]{4}) translates the scheduling decisions into GPU assignments and adjusts the parallelism topology.
The \underline{Controller}~(\dcircle[0.18]{5}) then delivers task and control messages to Workers, which execute denoising and report progress back to the Scheduler for next rounds.
\subsection{Intelligent Video Preemption}
\label{subsec:preemption}

\begin{figure}[t]
  \centering
  \includegraphics[width=\linewidth]{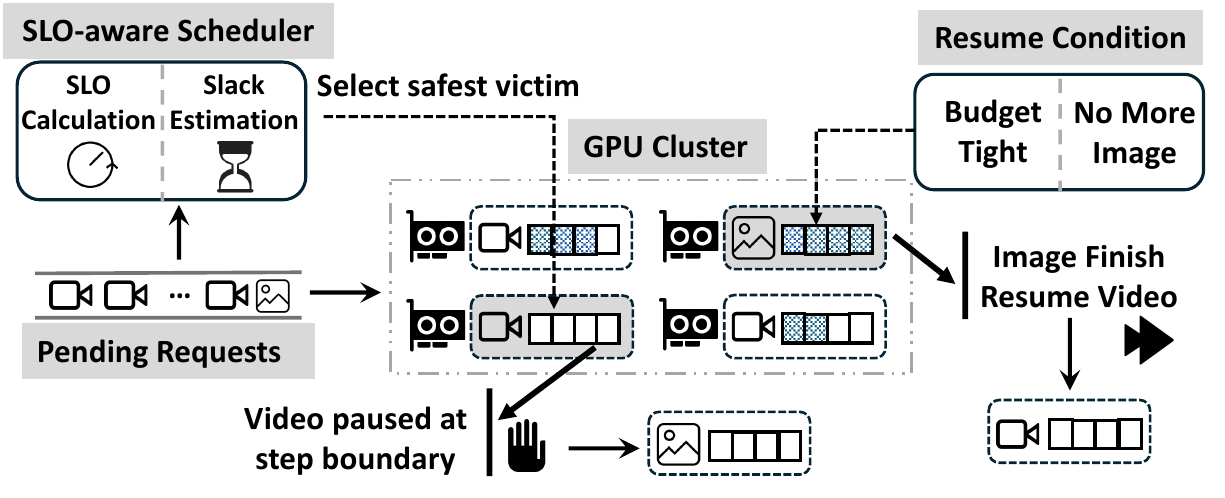}
  \caption{Illustration of intelligent video preemption. The scheduler computes the slack of running video jobs, pauses the safest one at a step boundary to serve an urgent image request, and resumes it when its deadline budget becomes tight, or no more image requests arrive.}
  \label{fig:sec_4-2_preemption-overview}
\end{figure}

In mixed serving, long-running video tasks can monopolize GPUs and block image requests (as discussed in \S\ref{sec: motivation}).
In this section, \Mname{} introduces the intelligent video preemption mechanism that pauses running video tasks at denoising step boundaries to free GPUs for urgent image requests.
The mechanism addresses two key questions: \emph{which} video to preempt, and \emph{when} to resume it, while ensuring that the preempted video can still meet its own deadline.
Figure~\ref{fig:sec_4-2_preemption-overview} illustrates the overall workflow.

\noindent \textbf{Potential Candidate Selection.}
When a latency-sensitive image request arrives and no GPU is available, the scheduler must select a running video to preempt. A naive strategy like preempting the most recently started video risks violating the candidate's deadline, while always preempting the longest-running video may unnecessarily interrupt a task close to completion, degrading the overall SLO attainment. 
Instead, \Mname{} leverages the predictability of per-step execution to compute the \emph{deadline slack} of every running video $v$:
\begin{equation} \small
\label{eq:slack}
\mathrm{slack}_v = D_v - t_{\mathrm{now}} - S_v^{\mathrm{rem}} \cdot T_{\mathrm{step}}(v),
\end{equation}
where $D_v$ is the video's deadline, $S_v^{\mathrm{rem}}$ is the number of remaining denoising steps, and $T_{\mathrm{step}}(v)$ is the offline profiled per-step latency obtained from Profiler under the current configuration.
Since $T_{\mathrm{step}}(v)$ is near-constant across runs (as discussed in \S\ref{sec: motivation}), the remaining runtime $S_v^{\mathrm{rem}} \cdot T_{\mathrm{step}}(v)$ can be estimated with high accuracy.

A positive slack means the video can tolerate a pause and still finish on time; a non-positive slack means any interruption would likely cause a deadline miss.
The scheduler ranks all running videos by descending slack and selects the highest-slack videos as preemption victims, since they have the most room to absorb the pause duration.
Videos with non-positive slack are excluded from consideration to avoid causing deadline violations.
The scheduler continues selecting victims until enough GPUs are freed to serve the pending image requests.

Beyond full preemption, the system can also release GPUs by downgrading the SP degree of running videos (e.g., reducing SP degree from 4 to 2 frees 2 GPUs) rather than pausing them entirely.
This provides a softer alternative that trades video latency for image capacity without fully interrupting the video's progress.
The optimal combination of preemption and SP reconfiguration is determined jointly by the SLO-aware scheduler (\S\ref{sec:slo-aware-scheduler}).

\noindent \textbf{Deadline-aware Resume Policy.}
Once the video is preempted, its latent state is retained in device memory, enabling near-instantaneous resumption without recomputation.
For this case, the key question is to decide when to resume the paused video. Resuming too early wastes the preemption opportunity, while resuming too late will miss the deadline.

To overcome this issue, \Mname{} triggers video resumption under two conditions:
The first is budget-tight: when the remaining time until the video's deadline falls below the estimated completion time, any further delay would cause a deadline miss, so the video must resume immediately. 
The second is idle: when no new image requests have arrived for a sustained period, the system resumes the paused video to maximize the utilization of the cluster. 
Together, these two conditions ensure that the video is paused for the maximum possible duration to serve image requests, while never risking its own SLO.
When the video resumes and additional GPUs are idle, the system can further upgrade its SP degree to accelerate the remaining steps, a decision coordinated by the SLO-aware scheduler (\S\ref{sec:slo-aware-scheduler}).

\begin{figure}[t]
  \centering
  \includegraphics[width=\linewidth]{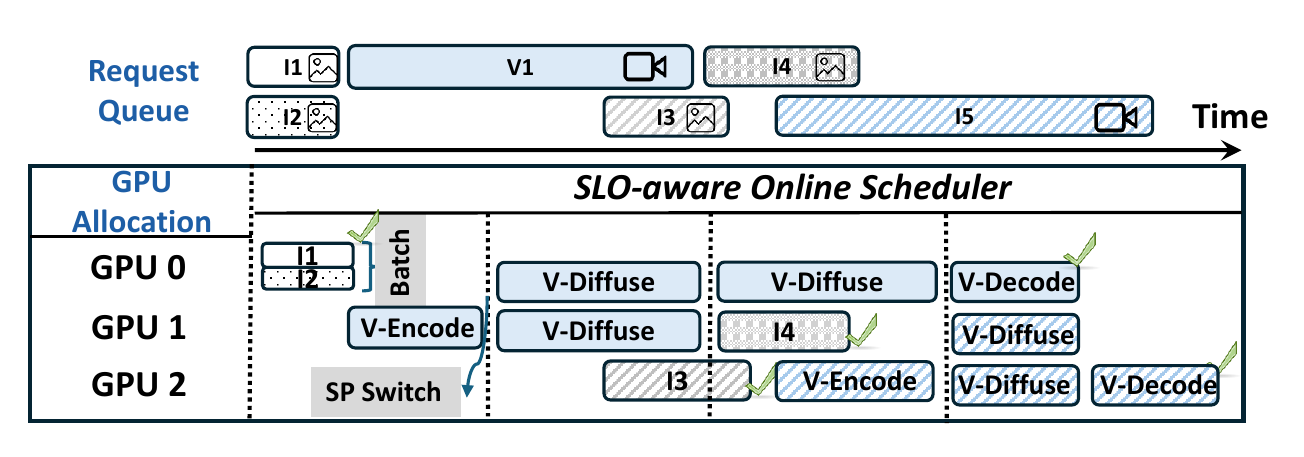}
  \caption{Adaptive resource allocation on the cluster serving mixed image and video requests. The SLO-aware online scheduler dynamically batches same-resolution images, adjusts the video SP degree at boundaries, and decouples stages on separate GPUs to maximize utilization.}
  \label{fig:sec_4-3_adaptive_allocation}
\end{figure}

\subsection{Adaptive Resource Allocation}
\label{subsec:Elastic-Resource-Reallocation}

Text-to-image and text-to-video workloads exhibit fundamentally different computational profiles,
a static resource configuration cannot efficiently serve if the workload mix changes over time.
In this section, \Mname{} addresses it through elastic resource allocation that operates along two complementary dimensions: elastic sequence parallelism for video tasks and SLO-aware dynamic batching for image tasks, continuously redistributing GPU capacity between modalities based on runtime demand (Figure~\ref{fig:sec_4-3_adaptive_allocation}).

\noindent \textbf{Elastic Sequence Parallelism.}
The typical video generation pipeline consists of two main stages: the DiT denoising stage and the VAE decoding stage.
As discussed in \S\ref{sec: motivation}, the DiT stage is compute-bound and occupies over 90\% of the total runtime, while the VAE stage is memory-bound and too lightweight to benefit from parallelism. Thus, \Mname{} efficiently decouples the two stages, letting the VAE stage always execute on a single GPU, while the DiT stage is parallelized via Sequence Parallelism across multiple GPUs.

Beyond efficient decoupling, \Mname{} supports runtime SP degree switching to tune the cluster conditions.
Since different video resolutions have different optimal SP degrees (\S\ref{sec: motivation}), the system must be able to self-evolve as the workload mix changes.
For example, when image requests arrive and GPUs are scarce, the system should downgrade a video's SP degree to free GPUs for image serving.
Conversely, when GPUs become idle, the system needs to upgrade the SP degree to accelerate video completion to meet the SLO.

\noindent \textbf{SLO-aware Dynamic Batching.}
Image requests that share the same resolution can be batched together to amortize GPU kernel launch overhead and improve throughput (as discussed in \S\ref{sec: motivation}).
However, naive batching can hurt SLO attainment: adding more requests to a batch increases the per-request latency, potentially causing earlier-arriving requests to miss their deadlines. 

In this part, \Mname{} implements deadline-aware batch formation. When an image request is dispatched to a GPU, the system attempts to build a batch by collecting additional same-resolution requests from the queue.
For each candidate, the system queries the Profiler to predict the end-to-end latency under the enlarged batch size and checks whether all requests in the batch can still meet their respective deadlines.
If adding the candidate would cause any request to violate its SLO, the candidate is deferred and the batch is deprecated.
To balance batching efficiency with latency, the system computes a dynamic wait budget based on the earliest deadline in the current batch: it waits for additional same-resolution requests to arrive only while the slack permits.
This allows the system to form larger batches under light load and dispatch smaller batches promptly under heavy load, adapting batch size to the real-time SLO pressure.

The two mechanisms are tightly coupled: GPUs freed by SP downgrade become available to serve batched image requests, and GPUs released after image batch completion can rejoin video SP groups.
The joint optimization of SP degree selection and image batching budget is handled by the SLO-aware scheduler (\S\ref{sec:slo-aware-scheduler}).

\subsection{SLO-aware Online Scheduler}
\label{sec:slo-aware-scheduler}
The scheduler jointly decides how many GPUs to reserve for image batching, which videos to continue or preempt, and when SP reconfiguration is worthwhile. We formulate it as an allocation problem solved by a lightweight DP algorithm. Table~\ref{tab:notation} summarizes the key notation.

\noindent \textbf{Problem Formulation.}
The system serves image requests $R_I$ and video requests $R_V$. Let $z_r \in \{0,1\}$ indicate whether request $r$ meets its deadline. The objective is to maximize deadline satisfaction:
\begin{equation} \small 
\label{eq:objective}
\max \sum_{r \in R_I \cup R_V} z_r .
\end{equation}

The scheduler is re-invoked at step boundaries and on scheduling events (e.g., new request arrivals), and $X_r(t)\subseteq\mathcal{G}$ denotes the allocation of request $r$ at round boundary $t$. Images use at most one GPU ($|X_i(t)|\in\{0,1\}$), while videos use $|X_v(t)|\in\{0\}\cup P$ GPUs. The allocation transition $X_r(t) \rightarrow X_r(t{+}\Delta_{\mathrm{round}})$ subsumes all scheduling actions, including start, resume, continue, preempt, and reconfigure in a single abstraction.
The next-round plan must satisfy the capacity constraint:
\begin{equation} \small 
\label{eq:capacity}
\sum_r |X_r(t+\Delta_{\mathrm{round}})| \le N .
\end{equation}

\begin{table}[t]
\small
\centering
\caption{Key notation in the \S\ref{sec:slo-aware-scheduler}.}
\vspace{-8pt}
\label{tab:notation}
\begin{tabular}{@{}cp{5.8cm}@{}}
\toprule
\textbf{Symbol} & \textbf{Description} \\
\midrule
$\mathcal{G}, N$ & GPU set $\{1,\dots,N\}$ and cluster size \\
$R_I, R_V$ & Image and video request sets \\
$D_r, A_r$ & Deadline and arrival time of request $r$ \\
$S_v^{\mathrm{rem}}(t)$ & Remaining steps of video $v$ at time $t$ \\
$X_r(t)$ & GPU alloc. of request $r$ at round boundary $t$ \\
$P$ & Set of valid SP degrees \\
$\Delta_{\mathrm{round}}$ & Scheduling round interval (in denoising steps) \\
$T_{\mathrm{img}}(b,w,h)$ & Profiled latency for image batch of size $b$ \\
$T_{\mathrm{step}}(w,h,F,p)$ & Profiled latency for one video denoising step \\
$\mathcal{C}_v(t)$ & Candidate set for video $v$ at time $t$ \\
$w(c), f(c)$ & GPU cost and score of candidate $c$ \\
$\ell_v(c,t)$ & Projected laxity of video candidate $c$ \\
\bottomrule
\end{tabular}
\end{table}

\noindent \textbf{Candidate Construction and Scoring.}
Directly optimizing over all GPU subsets is intractable. Instead, the scheduler constructs a small set of candidates that encode topology and placement constraints, scores them, and selects among them via DP. Algorithm~\ref{alg:dp-round-scheduler} outlines the three-stage procedure.

Stage~1 (lines~2--5) builds scored candidates for each request group. For images, since requests are single-GPU and benefit primarily from batching, we collapse scheduling into budgeted options: for each GPU budget $g\in\{0,\dots,N\}$, earliest-deadline-first (EDF) batching produces the best feasible plan $B^\star(g,t)$. The image score tracks two quantities: the number of deadline-satisfiable images $|B^\star(g,t)|$ (the \emph{recoverable count}), and a slack-based tiebreaker:
\begin{equation} \small
\label{eq:image-value}
f^{\mathrm{img}}_g(t)=
\sum_{i\in B^\star(g,t)}
\frac{1}{1+\max(0,\mathrm{slack}_i(g,t))},
\end{equation}
where $\mathrm{slack}_i(g,t)$ is the residual time between the predicted completion and the deadline.
For each video $v$, the scheduler generates a small anchored candidate set $\mathcal{C}_v(t)$, e.g., hold (pause), continue at the current SP, scale up or down, or resume on an available pool. Each candidate $c$ is scored by projected completion laxity $\ell_v(c,t) = D_v - \widehat{F}_v(c,t)$, where $\widehat{F}_v(c,t) = t + S_v^{\mathrm{rem}}(t) \cdot T_{\mathrm{step}}(w_v,h_v,F_v,p_v(c))$:
\begin{equation} \small
\label{eq:video-value}
f_v(c)=
\frac{1}{1+|\ell_v(c,t)|}, \quad
\mathrm{recoverable}(c) = \mathbf{1}[\ell_v(c,t)\ge 0].
\end{equation}
Each candidate also carries a boolean recoverable flag. A hold candidate carries zero value, allowing the solver to defer a video when images or urgent videos should take precedence.

\begin{algorithm}[t]
\small
\caption{SLO-aware DP Scheduler}
\label{alg:dp-round-scheduler}

\KwIn{round time $t$, GPU set $\mathcal{G}$, image set $\mathcal{I}(t)$, video set $\mathcal{V}(t)$, current allocations $\{X_r(t)\}$}
\KwOut{Allocation plan $\Pi^\star$}

$t^+ \gets t + \Delta_{\mathrm{round}}$\;

\tcc{\textcolor{blue}{Stage 1: Build scored candidates}}
\For{$g = 0$ \KwTo $N$}{
    $\mathcal{C}^{\mathrm{img}}_g(t) \gets \textsc{EdfBatch}(\mathcal{I}(t), g)$
    \tcp*[f]{\textcolor{blue}{Eq.~\eqref{eq:image-value}}}
}

\ForEach{$v \in \mathcal{V}(t)$}{
    $\mathcal{C}_v(t) \gets \textsc{GenVideoCandidates}(v,t,X_v(t),\mathcal{G})$
    \tcp*[f]{\textcolor{blue}{Eq.~\eqref{eq:video-value}}}
}

\tcc{\textcolor{blue}{Stage 2: Knapsack DP over video groups}}
Initialize $\mathrm{dp}[0][0] \gets (0, 0)$\;

\For{$j=1$ \KwTo $|\mathcal{V}(t)|$}{
    \For{$b=0$ \KwTo $N$
    }{
        $\mathrm{dp}[j][b] \gets (-\infty, -\infty)$\;

        \ForEach{$c \in \mathcal{C}_{v_j}(t)$}{
            \If{$w(c)\le b$ \textbf{and} \text{no GPU overlap with prev.\ selections}}{
                $s \gets \mathrm{dp}[j{-}1][b{-}w(c)] + \bigl(\mathrm{recoverable}(c),\; f(c)\bigr)$\;

                \lIf{$s > \mathrm{dp}[j][b]$}{
                    update $\mathrm{dp}[j][b]$, record backpointer
                }
            }
        }
    }
}

\tcc{\textcolor{blue}{Stage 3: Select best plan with img. eval.}}
\ForEach{terminal state $b$}{
    $g_{\mathrm{free}} \gets N - b$;
    combine video score with $\mathcal{C}^{\mathrm{img}}_{g_{\mathrm{free}}}(t)$\;
}
$(b^\star, \mathcal{S}^\star) \gets \textsc{BacktrackBestSolution}(\mathrm{dp})$\;
$\Pi^\star \gets \textsc{MaterializeAllocations}(\mathcal{S}^\star, \{X_r(t)\}, \mathcal{G})$\;

\Return $\Pi^\star$\;

\end{algorithm}

\noindent \textbf{Adaptive Online Selection.}
Stage~2 (lines~6--13) solves a knapsack-style DP over the video candidate groups. For each group $j$, the DP tries every candidate $c\in\mathcal{C}_{v_j}$ whose GPU set does not overlap with previously selected placements. Since each candidate is anchored to a concrete GPU set by the allocator, compatibility reduces to checking that no GPU is double-assigned. The DP tracks a lexicographic objective per state: the total recoverable count (number of candidates with $\ell_v \ge 0$), then the cumulative utility score $f(c)$ as a tiebreaker:
\begin{equation} \small
\label{eq:dp-transition}
\mathrm{dp}[j][b]
=
\max_{\substack{c \in \mathcal{C}_{v_j},\; w(c)\le b\\ \text{no GPU overlap}}}
\Bigl(
\mathrm{dp}[j-1][b-w(c)] + \bigl(\mathrm{recoverable}(c),\; f(c)\bigr)
\Bigr).
\end{equation}
After the video DP completes, the solver evaluates each terminal state: for the remaining free GPUs, it looks up the best image candidate $f^{\mathrm{img}}_g$ and selects the overall plan that maximizes the combined recoverable count across modalities.

Stage~3 (lines~14--17) backtracks to recover the optimal selection $\mathcal{S}^\star$ and calls \textsc{MaterializeAllocations} to convert it into concrete GPU assignments and runtime actions: hold, resume, continue, or reconfigure. The entire DP runs in $O(G \cdot N \cdot \bar{K})$ time, where $G{=}|\mathcal{V}(t)|$ video groups and $\bar{K}$ average candidates per group.

\section{Implementation}
We implement \Mname{} in about 10K LOC, which is built on Python. For diffusion model support, we integrate Hugging Face Diffusers~\cite{von-platen-etal-2022-diffusers} for pipeline management, and extend the WanPipeline~\cite{wan2025wan} with a preemptable wrapper that exposes step-level pause and resume through a persistent \texttt{VideoState} object (storing the intermediate latent tensor, prompt embeddings, and step index in GPU memory).
For image generation, we integrate StableDiffusion3~\cite{esser2024scaling} with configurable resolution and precision. Sequence parallelism is implemented on top of xFuser~\cite{fang2024xdit}, which provides Unified Sequence Parallel (USP) attention.

\section{Evaluation}
\label{sec:eval}

This section presents our evaluation settings and findings. It is organized into three parts: (1) end-to-end serving performance, comparing \Mname{} against baselines across SLO strictness, workload composition, arrival rate, and per-request latency; (2) ablation study, isolating the contribution of each system component and validating the co-serving design; and (3) sensitivity analysis on resolution heterogeneity and preemption overhead.

\subsection{Experimental Setup}

\noindent \underline{\textbf{Platform}}.
%
All experiments run on a single server equipped with 8$\times$NVIDIA RTX PRO 6000 Blackwell GPUs (96 GB GDDR7 each) connected via PCIe Gen5 ×16, running Ubuntu 24.04.4 LTS with CUDA 12.9, PyTorch 2.8, and NCCL for inter-GPU communication.  

\noindent \underline{\textbf{Models}}.
We evaluate two representative DiT-based diffusion models. For text-to-image, we use Stable Diffusion 3.5 Medium~\cite{esser2024scaling} (SD3.5 with 2.5B parameters), which supports resolutions from 256p to 1440p. For text-to-video, we use Wan2.2-T2V-5B~\cite{wan2025wan} with 5B parameters, generating videos with 41 or 81 frames across resolutions from 256p to 720p.

\noindent \underline{\textbf{Metrics}}.
Our primary metric is SLO attainment rate (SAR), defined as the fraction of requests completed before their deadlines, reported both overall and per request type.
Each request's deadline is set to $\sigma \times 1.5 \times$ its offline end-to-end latency~\cite{li2023alpaserve}, where $\sigma$ is an SLO scale factor that controls deadline tightness. We vary $\sigma$ from 0.8 (tight) to 1.3 (relaxed), with $\sigma{=}1.0$ as the default.
We also report per-request latency CDFs to characterize tail behavior.

%
\begin{figure}[t]
  \centering
  \includegraphics[width=\linewidth]{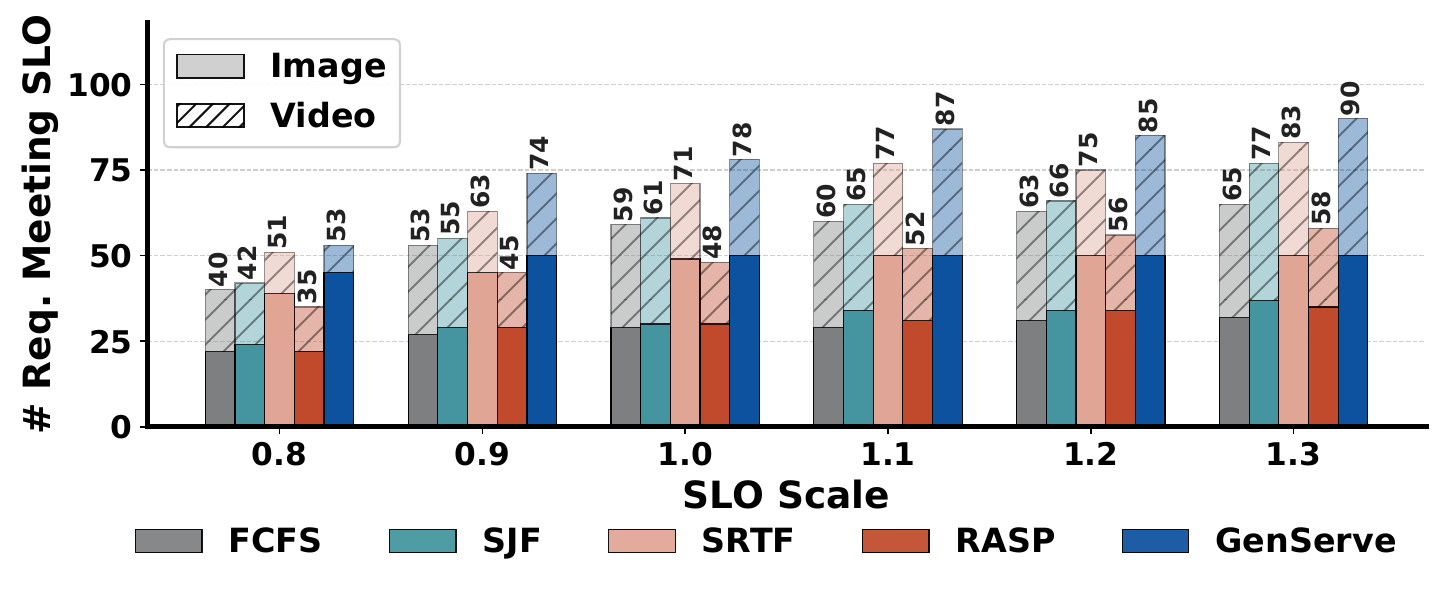}
  \caption{Number of requests meeting SLO versus SLO scale~$\sigma$ at the anchor configuration (balanced mix, 24\,req/min). Bars are split into image (solid) and video (hatched) contributions. \Mname{} leads at every $\sigma$, especially reaching 90 out of 100 at $\sigma{=}1.3$.}
  \label{fig:e2e_E1_slo_scale}
\end{figure}

\noindent \underline{\textbf{Datasets and Workload}}.
We synthesize mixed image and video workloads using real-world prompts:  DiffusionDB~\cite{wang2023diffusiondb} for image tasks and VBench~\cite{huang2024vbench} for video tasks.
We evaluate two arrival patterns: (1) Poisson, where requests arrive at a constant rate; and (2) Bursty, where requests concentrate in short intervals, creating load spikes that stress scheduling and resource allocation.
We vary workloads along three dimensions: (1) Task mix: the video-to-image ratio, including light (20:80), balanced (50:50), and heavy (80:20); (2) Request size: small, medium, and large, corresponding to image resolutions of $\{720p, 1024p, 1440p\}$ and video resolutions of $\{256p, 480p, 720p\}$; and (3) Resolution distribution: uniform or skewed (Dirichlet $\alpha{=}1.0$).
Each trace contains 100 requests, and each video request generates 81 frames.

\noindent \underline{\textbf{Baselines}}.
We compare \Mname{} against four baselines spanning common scheduling and allocation strategies:

\noindent \textbf{B1: FCFS}. First-come-first-served scheduling, where requests are served in arrival order.

\noindent \textbf{B2: SJF}. Shortest-job-first scheduling, where requests are ordered by estimated total runtime.

\noindent \textbf{B3: SRTF}. Shortest-remaining-time-first scheduling with preemption, where requests with the shortest remaining time have the highest priority.

\noindent \textbf{B4: RASP}. Resolution-aware sequence parallelism with static SP degree assignment, selecting SP degree $\{1, 2, 4\}$ for video resolution $\{256p, 480p, 720p\}$ based on Figure~\ref{fig: sec_3_video_stage_breakdown}.

\subsection{End-to-end Performance}
\label{sec:exp-e2e}
We evaluate SLO attainment rate (SAR) across SLO strictness, workload composition, system load, and per-request latency distributions.
Unless stated otherwise, the default configuration uses a balanced image-to-video mix (50:50), mean arrival rate of 24\,req/min, SLO scale $\sigma{=}1.0$, and uniform resolution distribution on 8 GPUs.

\noindent \textbf{E1: SLO Scale.}
Figure~\ref{fig:e2e_E1_slo_scale} shows overall, image, and video SAR as the SLO scale factor~$\sigma$ varies from 0.8 to 1.3.
\Mname{} achieves the highest overall SAR at every~$\sigma$: 78\% at $\sigma{=}1.0$, compared to 71\% for SRTF and 59\% for FCFS.
The gap widens at $\sigma{=}1.1$, where \Mname{} reaches 87\%---10\,pp above SRTF---because the moderate SLO slack provides just enough room for the DP solver to exploit both preemption and dynamic SP allocation.
At $\sigma{=}1.3$, \Mname{} achieves 90\% overall (100\% image, 80\% video), 7\,pp above SRTF (83\%).
Both SRTF and \Mname{} reach 100\% image SAR at $\sigma{\ge}1.0$ by interrupting long-running videos, while FCFS and SJF never exceed 74\%.
The key differentiator is video SAR: at $\sigma{=}1.0$, SRTF's aggressive preemption yields only 44\%, while \Mname{} achieves 56\%, because the DP solver limits preemption to cases where the net SLO gain is positive, avoiding cascading re-preemptions.
Under the tightest regime ($\sigma{=}0.8$), \Mname{} still leads overall by deliberately trading video SAR for image SAR---a decision the solver makes because image deadlines are the most constrained.

\begin{figure}[t]
  \centering
  \includegraphics[width=\linewidth]{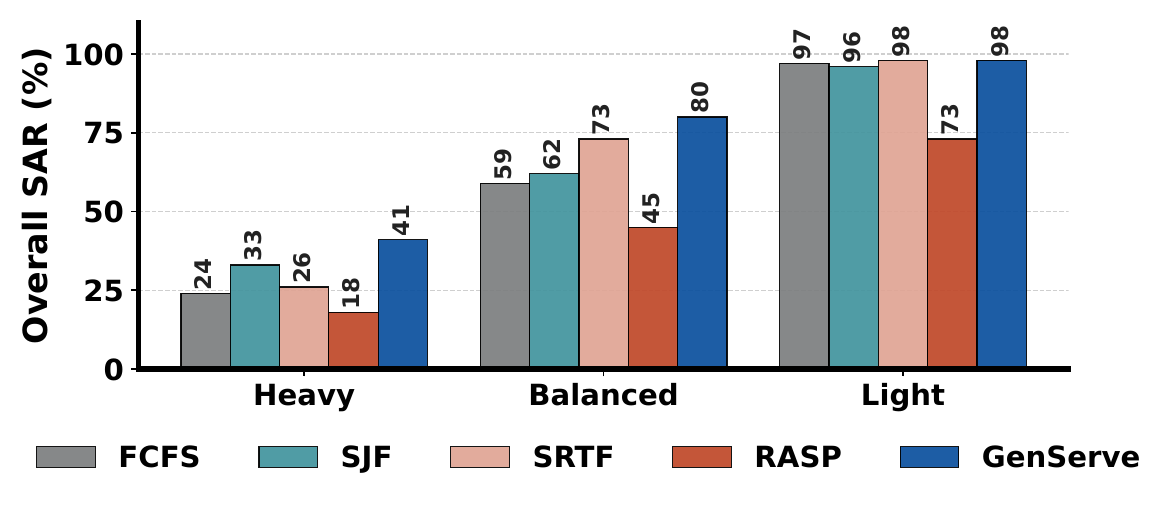}
  \caption{Overall SAR (\%) across workload mixes. \Mname{}'s advantage grows as the video proportion increases.}
  \label{fig:e2e_E2_workload_mix}
\end{figure}

\noindent \textbf{E2: Workload Mix.}
Figure~\ref{fig:e2e_E2_workload_mix} shows SAR across three image-to-video ratios at $\sigma{=}1.0$.
Under the image-dominated mix (Light), GPU contention is low: four of five methods achieve 96--98\% overall SAR, since the few video requests do not saturate the cluster.
RASP is the exception at 73\%, because its static SP allocation wastes parallelism on the few video requests.
At the balanced mix (Balanced), \Mname{} reaches 80\% overall, 7\,pp above SRTF (73\%).
The improvement stems from coordinating deadline-aware image batching with selective video preemption via the DP solver, which reduces average image wait from 3.6\,s (FCFS) to 0.1\,s.
Under the video-dominated mix (Heavy), all methods degrade sharply as 80 video requests compete for GPU time.
\Mname{} achieves 41\% overall, 8\,pp above SJF (33\%) and 15\,pp above SRTF (26\%).
\Mname{}'s video SAR of 31\% is $2.3{\times}$ that of SRTF (14\%): the DP solver avoids over-preemption by recognizing that, under heavy video load, repeatedly pausing videos harms more deadlines than it saves.

\begin{figure}[t]
  \centering
  \includegraphics[width=0.9\linewidth]{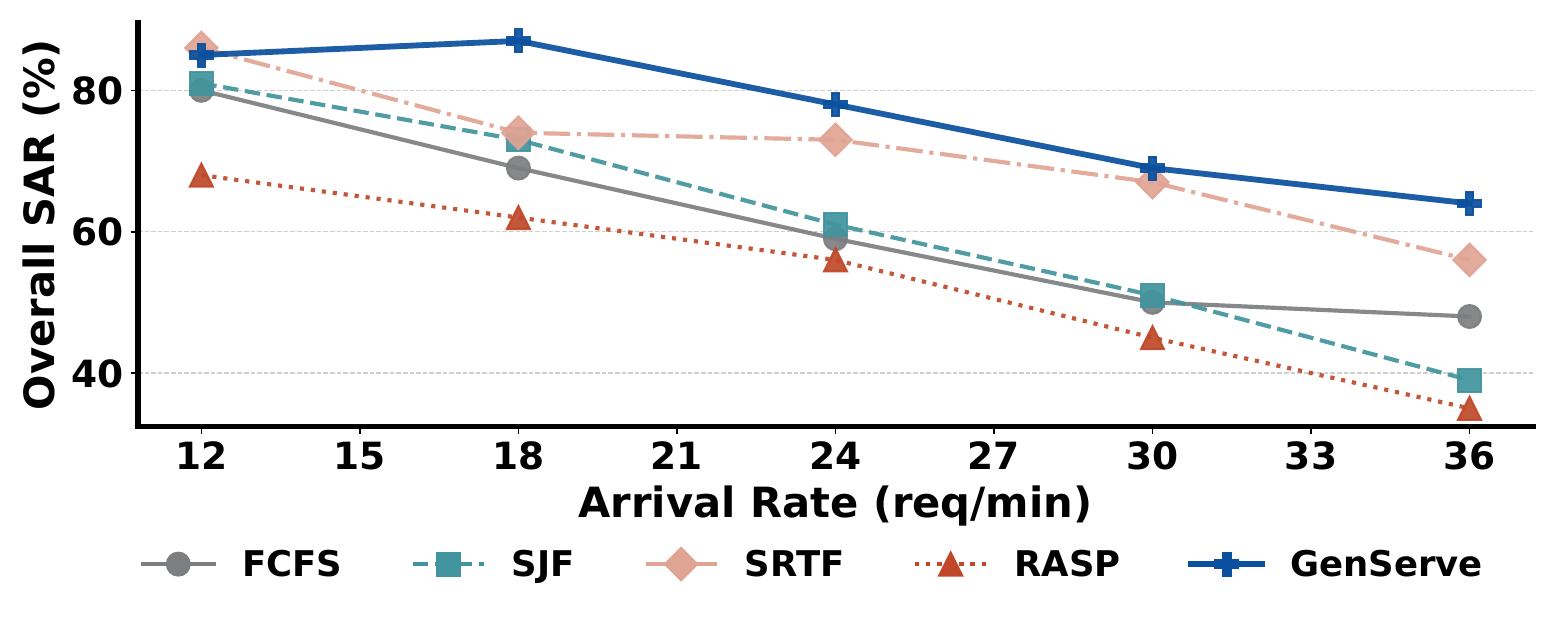}
  \caption{Overall SAR (\%) versus arrival rate (12--36\,req/min) at $\sigma{=}1.0$ and balanced mix. \Mname{} maintains the highest SAR across all load levels.}
  \label{fig:e2e_E3_arrival_rate}
\end{figure}

\noindent \textbf{E3: Arrival Rate.}
Figure~\ref{fig:e2e_E3_arrival_rate} shows SAR as the arrival rate increases from 12 to 36\,req/min.
At low load (12\,req/min), contention is minimal and preemption-based methods perform comparably (85--86\%); the DP solver provides little additional benefit when resources are abundant.
\Mname{}'s advantage emerges under moderate load: at 18\,req/min, it achieves 87\%, 13\,pp above SRTF (74\%).
At the highest load (36\,req/min), \Mname{} retains its lead at 64\% versus 56\% for SRTF, though the gap narrows as fundamental capacity limits dominate.
RASP degrades fastest because its fixed SP degrees cannot reallocate parallelism to absorb transient bursts.
SJF drops below FCFS at 36\,req/min (39\% vs.\ 48\%): under heavy load, the shortest-first policy starves long-running video requests whose deadlines expire while waiting.

\begin{figure}[t]
  \centering
  \includegraphics[width=0.9\linewidth]{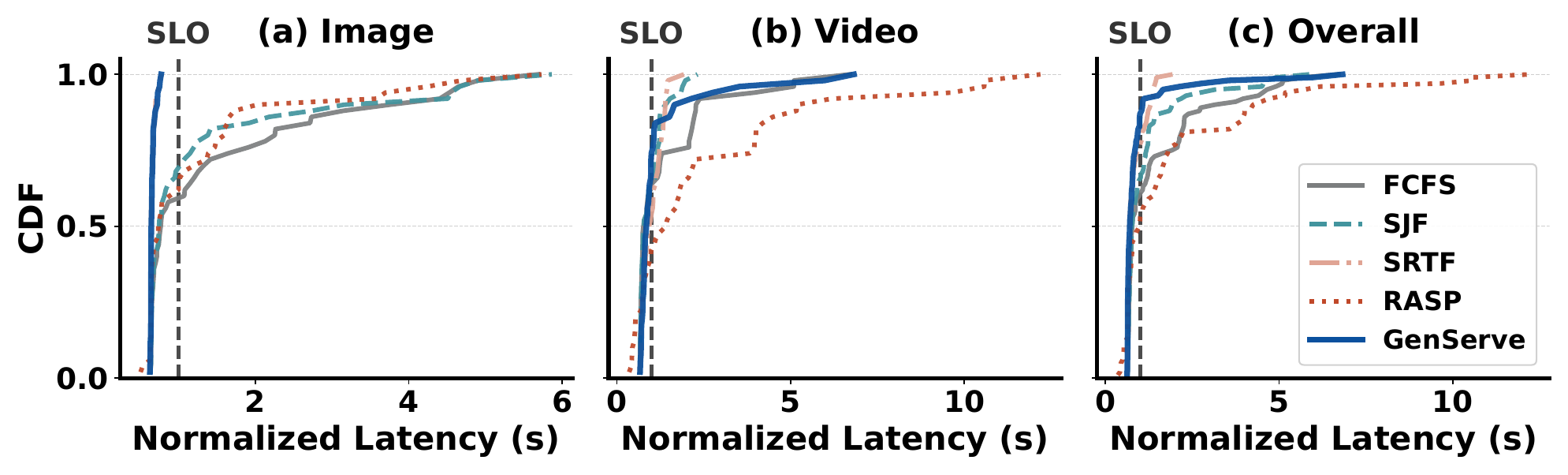}
  \caption{CDF of per-request turnaround latency at the default configuration ($\sigma{=}1.0$, balanced mix, 24\,req/min).}
  \label{fig:e2e_cdf}
\end{figure}

\noindent \textbf{E4: End-to-end Latency.}
Figure~\ref{fig:e2e_cdf} shows per-request turnaround latency CDFs at the default configuration.
For images, \Mname{} produces a steep CDF with p90 of 5.8\,s, a $3.1{\times}$ reduction over FCFS (18.0\,s).
Preemption eliminates the head-of-line blocking that causes FCFS and SJF images to queue behind long-running videos for up to 18\,s.
For videos, \Mname{} reduces median turnaround by 41\% (52\,s vs.\ 89\,s for FCFS) through dynamic SP, which accelerates per-step computation for high-resolution clips.
The video tail is wider (\Mname{} p99: 229\,s vs.\ FCFS: 166\,s): the longest-running videos accumulate multiple preemption cycles, each incurring state-save and context-switch overhead.
This is a deliberate trade-off: the DP solver prioritizes meeting SLOs for the majority of requests at the cost of extending the few that have already exceeded their deadline.

\subsection{Ablation Study}

\begin{figure}[t]
  \centering
  \includegraphics[width=0.9\linewidth]{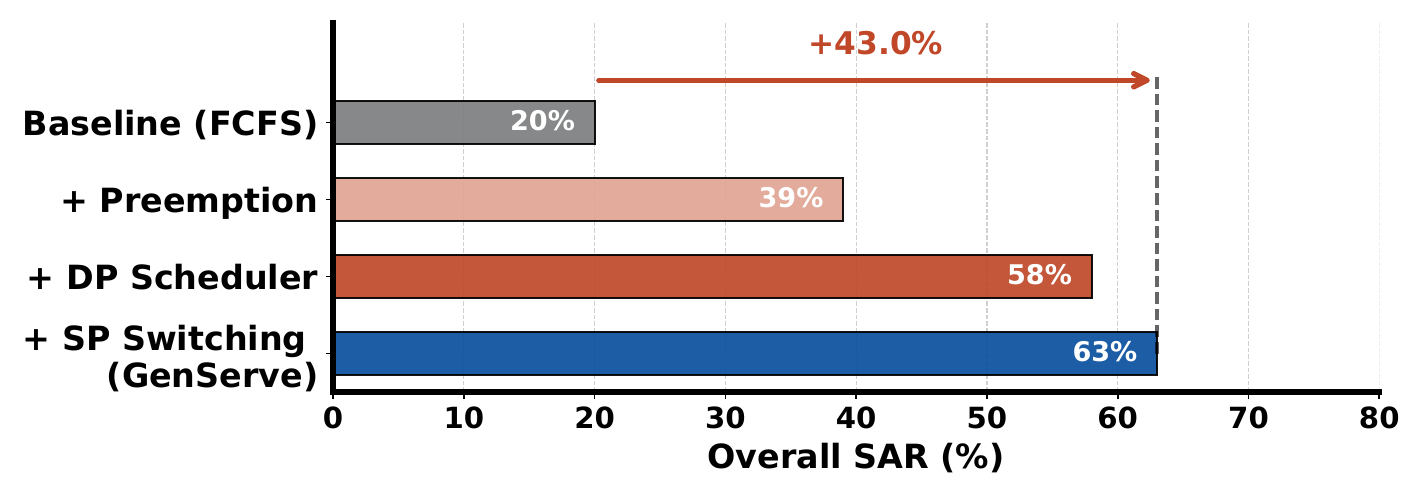}
  \caption{Ablation study under the skewed resolution setting ($\sigma{=}1.0$, balanced mix, 24\,req/min). Each bar group adds one component cumulatively.}
  \label{fig:e2e_E5_ablation}
\end{figure}

To isolate the contribution of individual system components, we incrementally enable each mechanism and measure the marginal SAR gain.
We use the skewed resolution setting ($\sigma{=}1.0$, balanced mix, Dirichlet $\alpha{=}1.0$), where high runtime variance amplifies the impact of scheduling decisions.

\noindent \textbf{Baseline (FCFS):}
Without preemption or dynamic resource management, FCFS achieves only 20\% overall SAR (20\% image, 20\% video).
Long-running high-resolution videos monopolize GPUs, causing images to wait an average of 15.9\,s, far exceeding their 2--8\,s deadlines.

\noindent \textbf{+ Preemption:}
Enabling video preemption lifts overall SAR to 39\%, with image SAR jumping to 68\%.
However, video SAR drops to 10\%: without a cost model, the scheduler preempts videos indiscriminately, and the resulting pause-resume overhead causes even more videos to miss their deadlines.
Preemption is necessary for image responsiveness but creates a zero-sum tradeoff when applied without coordination.

\noindent \textbf{+ DP Solver:}
Adding the DP solver on top of preemption yields another 19\,pp gain, reaching 58\% overall (94\% image, 22\% video).
The solver evaluates the net SLO impact before each preemption, reducing total preemptions and cutting average image wait from 4.6\,s to 0.2\,s.
Crucially, video SAR recovers from 10\% to 22\%: the solver avoids preempting videos that are close to completion, concentrating interruptions where the gain is largest.

\noindent \textbf{+ SP Switching (\Mname{}):}
Dynamic SP switching adds a final 5\,pp, bringing \Mname{} to 63\% overall (94\% image, 32\% video).
SP switching reduces average per-step video latency from 1154\,ms to 1121\,ms and provides the DP solver with an additional degree of freedom: instead of preempting a lagging video, the solver can accelerate it by upgrading its SP degree.
This further reduces preemptions while improving video SAR by 10\,pp.

These results yield three \underline{insights}:
(1)~preemption alone is the single most important mechanism for image SLO, but uncontrolled preemption damages video performance;
(2)~the DP solver is the critical coordinator, contributing an equal +19\,pp as preemption itself while lifting both image and video SAR;
(3)~SP switching expands the solver's action space beyond preempt-or-hold, enabling it to accelerate videos rather than interrupt them.

\begin{figure}[t]
  \centering
  \includegraphics[width=0.9\linewidth]{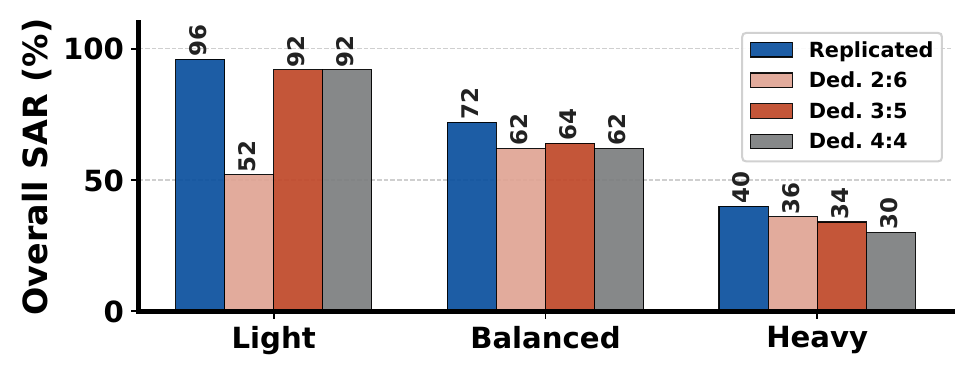}
  \caption{Dedicated GPU partitioning (\textit{X:Y} means $X$ image GPUs, $Y$ video GPUs) vs.\ replicated co-serving across workload mixes. The replicated design leads in all mixes by avoiding resource fragmentation.}
  \label{fig:e2e_E6_dedicated}
\end{figure}

\begin{table}[t]
    \small
    \centering
    \caption{SAR (\%) under uniform vs.\ skewed resolution distributions ($\sigma{=}1.0$, balanced mix, 24\,req/min). Best in \underline{\textbf{bold}}.}
    \resizebox{\linewidth}{!}{
    \begin{tabular}{l|ccc|ccc}
    \toprule
    \multirow{2}{*}{Method} & \multicolumn{3}{c|}{Uniform} & \multicolumn{3}{c}{Skewed} \\
    \cmidrule(lr){2-4} \cmidrule(lr){5-7}
     & Image & Video & Overall & Image & Video & Overall \\
    \midrule
    FCFS   & 58 & 60 & 59 & 20 & 20 & 20 \\
    SJF    & 60 & 62 & 61 & 18 & 30 & 24 \\
    SRTF   & 96 & 46 & 71 & 76 & 10 & 43 \\
    RASP   & 64 & 48 & 56 & 32 & 20 & 26 \\
    \rowcolor[HTML]{DAE8FC} \textbf{\Mname{}} & \underline{\textbf{100}} & \underline{\textbf{58}} & \underline{\textbf{79}} & \underline{\textbf{96}} & \underline{\textbf{28}} & \underline{\textbf{62}} \\
    \bottomrule
    \end{tabular}}
    \label{tab:e4_res_dist}
\end{table}

\noindent \textbf{Model Deployment.}
Figure~\ref{fig:e2e_E6_dedicated} compares static GPU partitioning (A: 2:6, B: 3:5, C: 4:4) against replicated co-serving across three workload mixes.
The replicated design leads in every mix: 40\% vs.\ 36\% (best dedicated) under heavy, 72\% vs.\ 64\% at balanced, and 96\% vs.\ 92\% under light load.
The core issue with dedicated partitioning is fragmentation: the optimal split depends on the workload mix, which shifts over time.
For example, Dedicated (A: 2:6) under image-heavy load achieves only 52\% because two GPUs cannot drain 40 image requests before their deadlines, while six video GPUs sit underutilized.
The replicated design avoids this by dynamically multiplexing all GPUs across both modalities.

\begin{table}[t]
    \footnotesize
    \centering
    \caption{DP solver wall-clock decision time versus number of concurrent request groups~$G$ ($N{=}8$ GPUs, $\bar{K}{\le}4$ candidates per group). Overhead is relative to a 720p base step (781\,ms). Statistics are aggregated across all E1 runs in \S\ref{sec:exp-e2e}.}.
    \resizebox{\linewidth}{!}{
    \begin{tabular}{crrr}
    \toprule
    Concurrent Groups ($G$) & Mean (ms) & Max (ms) & Overhead/Step (\%) \\
    \midrule
    1--2   & 0.30 & 0.7 & 0.09 \\
    3--4   & 0.24 & 1.0 & 0.13 \\
    5--6   & 0.25 & 1.5 & 0.19 \\
    7--8   & 0.31 & 1.8 & 0.23 \\
    9--12  & 0.41 & 1.9 & 0.24 \\
    \bottomrule
    \end{tabular}}
    \label{tab:dp_overhead}
\end{table}

\subsection{Sensitivity Analysis}
In this section, we evaluate how \Mname{} responds to key system parameters that impact scheduling efficiency and runtime overhead.

\noindent \textbf{Resolution Distribution.}
Table~\ref{tab:e4_res_dist} compares uniform and Dirichlet-skewed ($\alpha{=}1.0$) resolution distributions.
Skewed distributions concentrate requests at high resolutions, increasing average video runtime from 43\,s to 64\,s and amplifying SLO violations across all methods: FCFS drops from 59\% to 20\%, and SJF from 61\% to 24\%.
\Mname{} degrades from 79\% to 62\% (17\,pp drop), but widens its lead over the next-best method (SRTF, 43\%) from 8\,pp under uniform resolution to 19\,pp under skewed.
The widening gap arises because \Mname{}'s dynamic SP allocation adapts to the higher runtime variance: the DP solver assigns larger SP degrees to high-resolution videos, reducing their per-step latency, while static approaches cannot redistribute parallelism in response to the skewed request mix.


\noindent \textbf{DP Scheduler Overhead.}
Table~\ref{tab:dp_overhead} reports the DP solver's wall-clock decision time as a function of the number of concurrent request groups~$G$ (active videos plus one image group) on 8 GPUs.
The mean solve time grows from 0.30\,ms at $G{=}$1--2 to 0.46\,ms at $G{=}$9--12, consistent with the $O(G \cdot N \cdot \bar{K})$ complexity derived in \S\ref{sec:slo-aware-scheduler}.
Even in the worst observed case ($G{=}$9--12, max 1.9\,ms), the solver overhead is less than 0.25\% of a single 720p video step (781\,ms), confirming that the DP can be invoked at every scheduling round without becoming a bottleneck.

\noindent \textbf{Preemption Overhead.}
To support fine-grained scheduling, our preemption mechanism must incur a minimal performance penalty. Table~\ref{tab:preemption_scaling} evaluates this overhead across varying sequence parallelism (SP) degrees. The results indicate that pause latency is uniformly negligible ($\le 4.2\,\mu$s) across all configurations. While the resume overhead increases with the SP degree due to the coordination required across multiple GPUs—scaling from $0.036$\,ms at SP=1 to $0.868$\,ms at SP=8 for 720p. Preemption overhead remains a tiny fraction of the baseline step time ($<0.2\%$). This confirms that \Mname{} can afford aggressive preemption to optimize system-wide utility without degrading individual task throughput.

\noindent \textbf{Paused Video Memory Overhead.}
Table~\ref{tab:paused_memory} profiles the GPU memory footprint of each paused \texttt{VideoState}, which retains the latent tensor, denoising mask, and prompt embeddings on device.
The largest state (720p, 81 frames) consumes only 27.2\,MB.
At runtime, each GPU hosts at most one paused video at any time, because the scheduler resumes a paused video before pausing another on the same device.
Co-loading both pipelines occupies approximately 39\,GB per GPU, and peak memory during 720p inference reaches approximately 69\,GB, leaving over 26\,GB of headroom out of 96\,GB.
The paused state overhead ($<0.03\%$ of VRAM) is negligible and OOM is not a practical concern.

\noindent \textbf{SP Reconfiguration Overhead.}
All NCCL process groups for valid SP configurations are pre-created at startup, so switching SP degree is a pointer swap requiring no communicator re-initialization.
The paused video state resides entirely on the leader GPU without sharding, so no cross-GPU memory transfer occurs.
In practice, SP changes account for less than 0.4\% of total denoising steps across all experiments: the DP solver only triggers reconfiguration when the projected SLO benefit outweighs the transition cost.

\noindent \textbf{}
\begin{table}[t]
    \footnotesize
    \centering
    \caption{Preemption overhead summarized by 720p video with different sequence parallelism (SP) degrees. Results are averaged across 5 independent runs.}
    \resizebox{\linewidth}{!}{
    \begin{tabular}{llrrrr}
    \toprule
    Res. & SP & Base Step (ms) & Pause ($\mu$s) & Resume (ms) & Resume/Step (\%) \\
    \midrule
    \multirow{4}{*}{720p}
     & 1 & 781.57 & 3.40 & 0.036 & 0.005 \\
     & 2 & 780.16 & 3.05 & 0.464 & 0.060 \\
     & 4 & 780.66 & 3.00 & 0.592 & 0.076  \\
     & 8 & 781.15 & 3.05 & 0.868 & 0.112 \\
    \bottomrule
    \end{tabular}}
    \label{tab:preemption_scaling}
\end{table}

\begin{table}[t]
    \footnotesize
    \centering
    \caption{GPU memory footprint of a single paused \texttt{VideoState} (81 frames, SP=1). Latent and mask tensors are float32; prompt embeddings are bfloat16.}
    \begin{tabular}{lrrrr}
    \toprule
    Resolution & Latent (MB) & Mask (MB) & Embeds (MB) & Total (MB) \\
    \midrule
    256p & 1.3  & 1.3  & 3.5 & 6.2  \\
    480p & 4.6  & 4.6  & 3.5 & 12.8 \\
    720p & 11.8 & 11.8 & 3.5 & 27.2 \\
    \bottomrule
    \end{tabular}
    \label{tab:paused_memory}
\end{table}
\section{Related Work}

\subsection{Diffusion Model Serving Systems}
\textbf{Parallelism and Scheduling.} The quadratic attention complexity of DiTs has made sequence parallelism (SP) indispensable for distributed inference. StreamFusion~\cite{yang2026streamfusion} and xDiT~\cite{fang2024xdit} introduce topology-aware and hybrid SP strategies to reduce communication overhead. On the scheduling side, TridentServe~\cite{xia2025tridentserve} decomposes diffusion pipelines into coarse-grained stages for flexible GPU allocation, while TetriServe~\cite{lu2026tetriserve} models serving as a step-level problem with dynamic SP adjustment.
These systems target homogeneous workloads; \Mname{} extends step-level scheduling to heterogeneous T2I/T2V co-serving with joint preemption, batching, and SP optimization.

\noindent \textbf{Caching.} To accelerate iterative denoising, MixFusion~\cite{sun2026mixfusion}, NIVARA~\cite{agarwal2023approximate}, and FlexCache~\cite{sun2024flexcache} cache and reuse intermediate feature maps across denoising steps to skip redundant computation.
These techniques are orthogonal to \Mname{}'s scheduling optimizations and can be integrated to further reduce per-step cost.

\noindent  \textbf{Multi-Model Routing.} DiffServe~\cite{sun2024flexcache} constructs model cascades that route queries to lightweight or heavyweight models based on difficulty, while MoDM~\cite{xia2026modm} directs requests across experts to balance cost and quality.
These are complementary to our intra-model resource management.

\subsection{Hybrid Serving Systems}

\textbf{Disaggregation and Stage-Aware Scaling.} vLLM-Omni~\cite{yin2026vllm} introduces full Encoder-Prefill-Decode disaggregation for any-to-any models, while EPD-Serve~\cite{bai2026epd} enables independent stage scaling to optimize cross-modal throughput. These systems target multimodal LLM pipelines where the bottleneck shifts across stages.
\Mname{} differs by targeting pure diffusion workloads, exploiting predictable per-step cost and step-boundary preemptibility for SLO-driven resource adaptation.

\noindent  \textbf{Scheduling and Resource Allocation.} ModServe~\cite{qiu2025modserve} employs modality-aware resource pools, while Cornserve~\cite{chung2026cornserve} introduces a cell-based runtime for dynamic routing by modality.
Unlike these partitioned designs, \Mname{} adopts replicated co-serving where all GPUs dynamically serve both T2I and T2V, avoiding fragmentation (\S\ref{sec:eval}).

\section{Conclusion}
In this work, we presented \Mname{}, a heterogeneity-aware co-serving system that leverages the inherent predictability and step-level preemptibility of diffusion models to maximize SLO attainment for mixed image and video workloads. \Mname{} combines intelligent preemption, elastic sequence parallelism, and dynamic batching with an SLO-aware DP-based scheduler that jointly coordinates resource allocation across all concurrent requests. Our evaluation demonstrates that these design choices yield substantial gains in SLO attainment across diverse workload configurations. More broadly, this work highlights how exposing and exploiting structural properties of diffusion workloads can unlock new opportunities for efficient large-scale model serving.

\bibliographystyle{plain}
\bibliography{refs}

\end{document}